\documentclass{aa}  

\usepackage{graphicx}
\usepackage{subcaption}
\usepackage{float}
\usepackage{placeins}
\usepackage{txfonts}
\usepackage{hyperref}
\usepackage{cleveref}
\crefname{figure}{Fig.}{Figs.}
\crefname{subfigure}{Fig.}{Figs.}
\crefname{equation}{Eq.}{Eqs.}
\crefname{table}{Table}{Tables}
\crefname{section}{Sect.}{Sects.}
\crefname{subsection}{Sect.}{Sects.}

\usepackage{siunitx}
\DeclareSIUnit{\pix}{pix}
\newcommand{\planet}{HD~209458\,b}

\usepackage{natbib}
\usepackage{lastpage}

\begin{document} 

   \title{Osiris revisited: Confirming a solar metallicity and low C/O in \planet}

   \author{N. Bachmann\inst{1,2}
          \and
          L. Kreidberg\inst{1}
          \and
          P. Molli\`ere\inst{1}         
          \and
          D. Deming\inst{3} 
          \and
          S.-M. Tsai\inst{4,5}}

   \institute{Max-Planck-Institut für Astronomie, Königstuhl 17, 69117 Heidelberg, Germany
              \and 
              Ruprecht-Karls-Universität Heidelberg, Fakultät für Physik und Astronomie, Im Neuenheimer Feld 226, 69120 Heidelberg, Germany\\
              \email{bachmann@mpia.de}
              \and
              Department of Astronomy, University of Maryland at College Park, MD 20742, USA
              \and
              Department of Earth and Planetary Sciences, University of California, Riverside, CA, USA
              \and
              Institute of Astronomy \& Astrophysics, Academia Sinica, Taipei 10617, Taiwan}

   \date{Received May, 19 2025; accepted June, 18 2025}

  \abstract{\planet{} is the prototypical hot Jupiter and one of the best targets available for precise atmosphere characterisation. Now that spectra from both Hubble Space Telescope (HST) and James Webb Space Telescope (JWST) are available, we can reveal the atmospheric properties in unprecedented detail. In this study, we perform a new data reduction and analysis of the original HST/WFC3 spectrum, accounting for the wavelength dependence of the instrument systematics that was not considered in previous analyses. This allows us to precisely and robustly measure the much-debated $\mathrm{H_2O}$ abundance in \planet's atmosphere. We combine the newly reduced spectrum with archival JWST/NIRCam data and run free chemistry atmospheric retrievals over the $\SI{1.0}{} - \SI{5.1}{\micro\metre}$ wavelength range, covering possible features of multiple absorbing species, including $\mathrm{CO_2}$, $\mathrm{CO}$, $\mathrm{CH_4}$, $\mathrm{NH_3}$, $\mathrm{HCN}$, $\mathrm{Na}$, $\mathrm{SO_2}$, and $\mathrm{H_2S}$. We detect $\mathrm{H_2O}$ and $\mathrm{CO_2}$ robustly at above $\SI{7}{\sigma}$ significance, and find a $\SI{3.6}{\sigma}$ preference for cloudy models compared to a clear atmosphere. For all other absorbers we tested, only upper limits of abundance can be measured. We use Bayesian model averaging to account for a range of different assumptions about the cloud properties, resulting in a water volume mixing ratio of $0.95^{+0.35}_{-0.17} \:\times$ solar and a carbon dioxide abundance of $0.94^{+0.16}_{-0.09} \:\times$ solar. Both results are consistent with solar values and comparable to predictions from the \texttt{VULCAN} 1D photochemistry model. Combining these values with a prior on the $\mathrm{CO}$ abundance from ground-based measurements, we derive an overall atmospheric composition comparable to solar metallicity of $\mathrm{[M/H]} = 0.10^{+0.41}_{-0.40}$ and very low C/O of $0.054^{+0.080}_{-0.034}$ with a $\SI{3}{\sigma}$ upper limit of $0.454$. This indicates a strong enrichment in oxygen and depletion in carbon during \planet's formation.}


   \maketitle
%
%--------------------------------------------------------------------
\section{Introduction}
\planet{} (also named Osiris) is one of the best characterised and most extensively studied exoplanets. Initially discovered through radial velocity measurements in 1999 \citep{Henry_2000}, it was the first exoplanet observed transiting its host star \citep{Charbonneau_2000, Henry_2000}. Since then, it has become a prime target for transit observations \citep[e.g.][]{Deming_2013, Madhusudhan_2014, MacDonald_2017, Giacobbe_2021, Xue_2023} due to its relatively bright host star and favourable planet-to-star radius ratio. \planet{} was the first exoplanet for which an atmosphere was observed \citep{Charbonneau_2002} (though the detection was later called into question by ground-based observations \citet{Casasayas-Barris_2020}). An undisputed detection of the atmosphere came from HST/WFC3 spatial scan observations, which revealed a strong signature of water vapour \citep{Deming_2013}. Water is among the easiest molecules to detect and analyse in exoplanetary atmospheres because of the numerous observable ro-vibrational bands in the near and mid-infrared, now accessible to multiple instruments. Additionally, water remains a key tracer of planet formation theories \citep{Öberg_2011, Helled_2014, Sing_2016, Öberg_2016}. Since the first atmospheric observations of \planet, clouds have been inferred in the transmission spectrum \citep{Charbonneau_2002, Deming_2013} to account for the low abundances measured.

The water abundance in the atmosphere of \planet, as measured using retrieval techniques, has been a topic of ongoing debate. Earlier studies reported sub-solar water abundances ($<-4.5$ in log volume mixing ratio) \citep{Madhusudhan_2014, MacDonald_2017, Brogi_2017, Welbanks_2019, Pinhas_2019}, therefore contradicting most planetary formation models, which predict solar or higher oxygen and water abundances based on core accretion scenarios \citep{fortneymordasini2013, Öberg_2016, Booth_2017}. Potential explanations for the low water abundance include high-altitude opaque clouds or hazes, although such factors were considered in some of the analyses \citep{MacDonald_2017}. These studies were primarily based on the transmission spectrum observed with the Hubble Space Telescope's (HST) Wide Field Camera 3 (WFC3) \citep{Deming_2013}. However, more recent work has suggested higher water abundances, with some studies reporting solar to super-solar values based on the same HST data \citep{Line_2016, Sing_2016, Tsiaras_2018, Welbanks_2021}. 

Additionally, several molecular absorbers, including sodium ($\mathrm{Na}$) \citep{Charbonneau_2002}, methane ($\mathrm{CH_4}$), carbon monoxide ($\mathrm{CO}$), carbon dioxide ($\mathrm{CO_2}$) \citep{Swain_2009, Madhusudhan_2009, Snellen_2010}, ammonia ($\mathrm{NH_3}$), and/or hydrogen cyanide ($\mathrm{HCN}$) \citep{MacDonald_2017, Giacobbe_2021}, have claimed findings in the planet's transmission spectrum. The simultaneous detection of $\mathrm{H_2O}$ and $\mathrm{HCN}$ of \citet{Giacobbe_2021} led to more discussion as it suggests a super-solar C/O ($\approx 1$), but it could not be confirmed in later studies by \citet{Blain_2024} and \citet{Xue_2023}.

Following the launch of the James Webb Space Telescope (JWST) in December 2021, its high precision and resolution have made it a powerful tool for exoplanetary and atmospheric studies. JWST's enhanced capabilities enable unprecedented spectral resolution, opening new possibilities for detecting molecules in exoplanetary atmospheres. Given its high signal-to-noise ratio (S/N) for transmission spectroscopy, \planet{} was one of the first targets observed with JWST's NIRCam instrument. \citet{Xue_2023} published the NIRCam transmission spectrum of \planet, along with an equilibrium chemistry retrieval analysis and a joint retrieval of their new spectrum with archival HST/WFC3 data. Their results support the presence of higher water abundances in the planet's atmosphere \citep{Xue_2023} due to the enrichment in oxygen supported by the low C/O and the comparable to solar metallicity.

Most previous studies relied on the HST transmission spectrum published by \citet{Deming_2013}, which was reduced using a different method than the one commonly used today. Therefore, for this study, we re-reduced the HST/WFC3 spectral data using the \texttt{PACMAN} pipeline \citep{Zieba_2022}, with particular attention to instrument systematics. We compared two different approaches to address these. To investigate the previously reported low water abundances, we performed atmospheric retrievals on our newly reduced HST transmission spectrum, both individually and in combination with the JWST/NIRCam data \citep{Xue_2023}, using \texttt{petitRADTRANS} \citep{Molliere_2019, Nasedkin_2024, Blain_2024_pRT}. 

The structure of the paper is as follows: In \cref{sec:data_red}, we describe the data reduction process using the \texttt{PACMAN} pipeline \citep{Zieba_2022}. Section~\ref{sec:HST_retrievals} presents the setup and results of our atmospheric retrieval analysis based on the newly reduced HST transmission spectrum, using \texttt{petitRADTRANS}. In \cref{sec:comb_retrievals}, we discuss the results of our analysis of the combined HST and JWST spectra, including the calculation of the metallicity and C/O from the free-chemistry retrieval results, and a comparison with predictions from the \texttt{VULCAN} 1D model \citep{Tsai_2021}. We summarise the key findings in \cref{sec:conclusions}.
%--------------------------------------------------------------------
\section{Data reduction and light curve fits}
\label{sec:data_red}
In this section we describe our new reduction of the HST/WFC3 spectroscopy data of \planet.

\subsection{Data reduction with \texttt{PACMAN}}
\label{subsec:data_red}
\begin{figure*}
   \centering
   \includegraphics[scale=1.0]{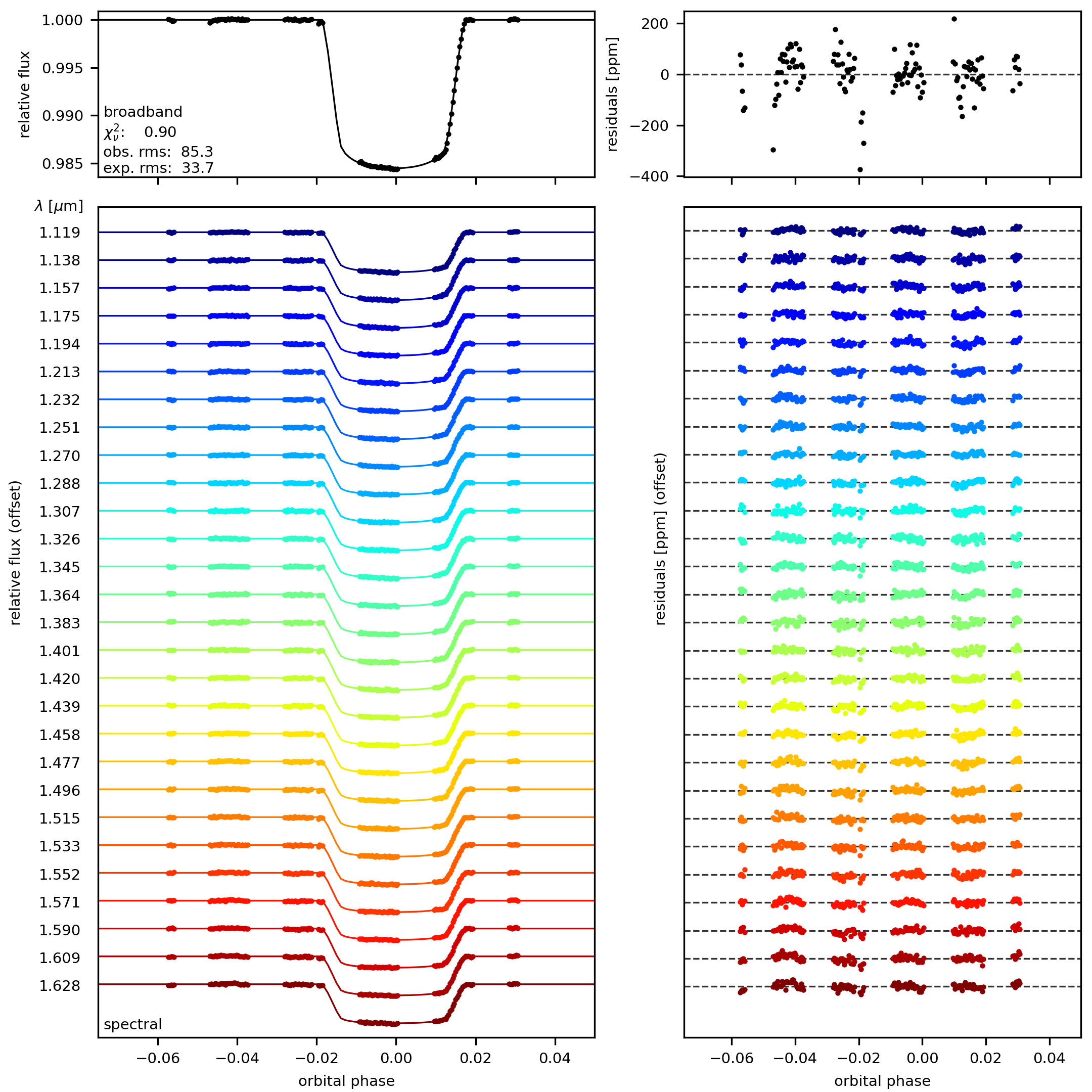}
      \caption{White light curve of \planet's HST/WFC3 transit with the best-fit transit model (upper left panel) and residuals of the fit (upper right panel), as well as the spectral light curves (lower left panel; offset) with fit transit models and the corresponding residuals (lower right panel).}
         \label{fig:wlc_fit}
\end{figure*}

We re-analysed the data from the first HST/WFC3 transmission spectroscopy observation of \planet. These observations used the WFC3/G141 grism \citep{Deming_2013}. The instrument was operated in spatial scan mode to account for the relatively bright host star ($\mathrm{m_J} = \SI{6.59}{mag}$\footnote{\url{https://simbad.cds.unistra.fr/simbad/sim-id?Ident=HD+209458}, 15.11.2023}) and ensure individual pixels did not saturate. As this was a first demonstration of spatial scan observations, there is only one scan direction, in contrast to the two directions typically used in more recent measurements. The data set contains one visit with five orbits with a varying number of scans per orbit, resulting in a total of 125 exposures, each lasting $\SI{32.9}{\second}$. The grism provides a wavelength coverage from $\SI{1.0}{\micro\metre}$ to $\SI{1.7}{\micro\metre}$. The breaks visible in the data in \cref{fig:wlc_fit} are not coincident with the orbits but occur due to scan setups and breaks for writing the data to memory. Since this data set is one of the first spatial scan observations, setups and data storage took some time and could not be controlled in the observation scheduling, as the programme was not originally designed for spatial scan mode.

For the majority of the data reduction, we used the \texttt{PACMAN} pipeline \citep{Zieba_2022}, which uses the \_ima.fits files provided by the data archive. After applying a barycentric correction to account for HST's orbital motion, a reference spectrum for the wavelength calibration is computed by multiplying a stellar model from the Kurucz stellar atmosphere atlas \citep{Kurucz_1993} with the grism bandpass. For this stage of \texttt{PACMAN}, we used a stellar effective temperature of $\SI{6030}{\kelvin}$, a stellar surface gravity of $\mathrm{log(g)} = 4.31$, and a stellar metallicity of $-0.0055$ \citep{Rosenthal_2021}. The star's position is determined from the direct image, and this information is used to extract the spectrum for each exposure.

To convert the 2D spatial scan spectral images from HST into a 1D spectrum, \texttt{PACMAN} employs the optimal extraction method \citep{Horne_1986}. To estimate and subtract the background, we set a threshold at $\SI{1000} \ {\rm electrons \ s^{-1}}$, considering all pixels with lower flux as background. Outliers greater than $\SI{15}{\sigma}$ were masked in the optimal extraction. The wavelengths are calibrated using the aforementioned template spectrum, effectively correcting for a shift of the extracted spectrum across exposures. The data are then binned into 28 wavelength bins using the same ones as in \citet{Deming_2013} for comparison. For the white light curve fit, a broadband light curve was generated by integrating over all wavelengths.

Finally, we fit the white light curve and spectroscopic light curves using a \texttt{batman} model for the astrophysical signal \citep{Kreidberg_2015} and several different models for the instrument systematics (described below). In our analysis, we included all orbits, omitting only the first five exposures of the first orbit, following \citet{Deming_2013}, because they exhibit the strongest ramp behaviour. Often, the first orbit is omitted altogether to avoid more pronounced instrument systematics. In our case, this was not necessary since only the first exposures showed a very strong systematic behaviour. In addition, we wanted to directly compare our new data reduction with the original one from \citet{Deming_2013}.

We adopted the orbital period of $\SI{3.52474859}{days}$ from \citet{Bonomo_2017} and set the eccentricity to $0.01$ \citep{Rosenthal_2021}. Since transit spectroscopy is not very sensitive to it, an eccentricity of zero would not change the results significantly. To model the effect of instrument systematics, which influence the shape of the light curve flux over time, we used exponential ramps at the beginning of each orbit (\cref{eq:ramp}) and a visit-long second-order polynomial along with a constant. Figure~\ref{fig:wlc_fit} shows the white light curve fit, including the new method explained in \cref{subsec:Deming_method}, and \cref{fig:wlc_corner} shows the corner plot with the fit parameters. Some of the best-fit parameters from the white light curve fit were fixed for the spectroscopic light curve fits. These include the transit midtime $\mathrm{t_0} = \SI{2456196.29}{BJD}$ , the semi-major axis $\mathrm{a} = \SI{8.79}{R_*}$, and the inclination $\mathrm{i} = \SI{86.7}{\degree}$. We calculated the limb darkening parameters using the \texttt{ExoTiC-LD} code \citep{Grant_2024} and fixed them during the fits. 

\texttt{PACMAN} uses the MCMC algorithm \texttt{emcee} \citep{Foreman_2013} to estimate the parameters. In each wavelength bin, the posterior distribution for each parameter is sampled using the uniform priors given in \cref{tab:slc_params}, 30 walkers, and 4000 steps (with 2000 as burn-in). Convergence is ensured by running the sampler for at least 50 times the autocorrelation time. To check for red noise, we plot the standard deviation (rms) of the residuals versus the bin size for each wavelength bin. With this we can ensure that the rms follows the inverse square root law expected for Gaussian uncorrelated noise.

We tested two different models to account for the instrument systematics in the spectroscopic light curves. The first one uses an exponential function (\cref{eq:ramp}) at the beginning of each orbit to account for the ramp in the data, similar to the white light curve fit.
\begin{equation}
    f(r_1, r_2, t_{\rm orb}) = 1 - e^{(-r_1 \cdot t_{\rm orb} - r_2)}
    \label{eq:ramp}
\end{equation}
Here, $r_1$ and $r_2$ are free parameters (see \cref{tab:slc_params}) and $t_{\rm orb}$ is the time in the observed orbit. This model will be referred to as "model\_ramp" in the following, and it accounts for the fact that the amplitude of the instrument systematics is dependent on the level of illumination, which in turn is wavelength dependent. In addition to the exponential ramps, a visit-long linear slope is fit to the data. The second model, "divide\_white", assumes that the systematic parameters of the spectroscopic light curves are wavelength independent and share the same shape as those of the white light curve. Thus, these systematics can be taken into account by using the systematics model of the white light curve \citep[cf. Eqs. 2 and 3 in ][]{Kreidberg_2014}. \cref{tab:slc_params} summarises the spectroscopic light curve fit parameters. 

\begin{table}[!htb]
\centering
\caption{Parameters for the spectroscopic light curve fits with assigned uniform priors ($U$).}
\begin{tabular}{c l}
 \hline
 \multicolumn{1}{c}{parameter} & \multicolumn{1}{c}{prior} \\
 \hline \hline
 $R_{pl} \: [\SI{}{R_*}]$ & $U\,(-0.01, 0.4)$ \\
 constant $c$ & $U\,(-8.0, 12.0)$ \\
 linear slope $v$ & $U\,(-0.5, 0.5)$ \\
 $r_1$ & $U\,(-1.0, 5.0)$\\
 $r_2$ & $U\,(-1.0, 50.0)$\\
 \hline
\end{tabular}
\tablefoot{The exponential ramp parameters $r_1$ and $r_2$ are only used for the "model\_ramp" method. The constant $c$ is given in log$_{10}$, so that the average flux is given by $10^{c}$.}
\label{tab:slc_params}
\end{table}

\begin{figure*}[!htb]
\centering
     \begin{subfigure}[b]{0.497\textwidth}
         \centering
         \includegraphics[width=1.0\linewidth]{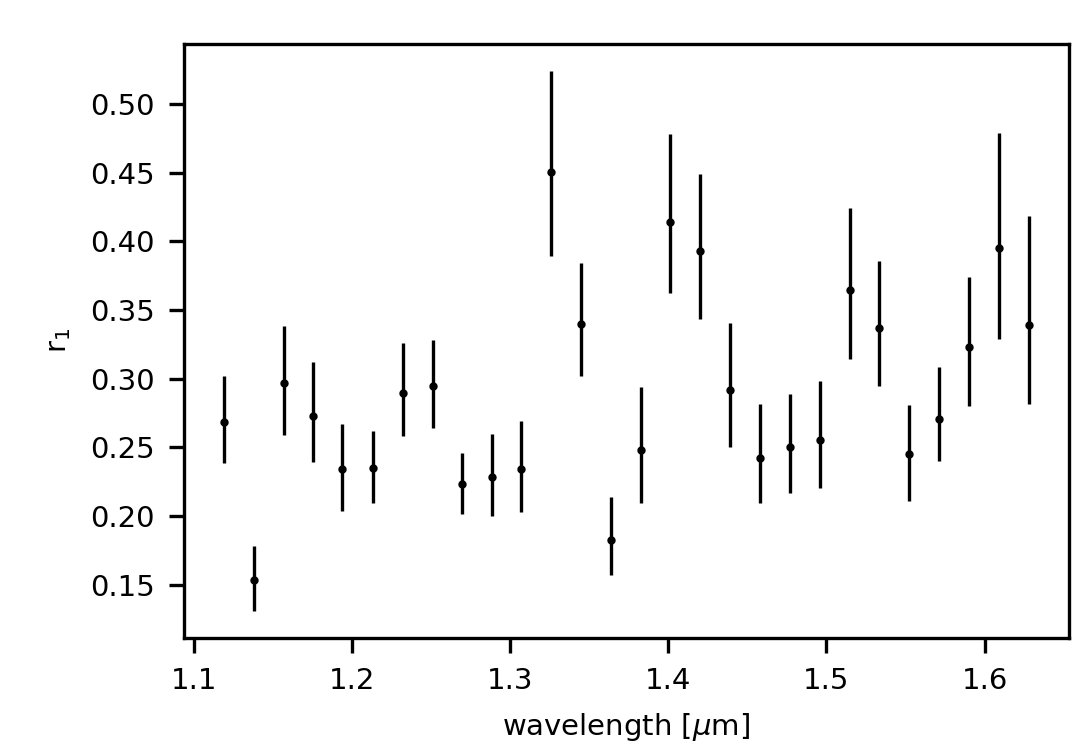}
         \label{fig:r1}
     \end{subfigure}
     \hfill
     \begin{subfigure}[b]{0.497\textwidth}
         \centering
         \includegraphics[width=1.0\linewidth]{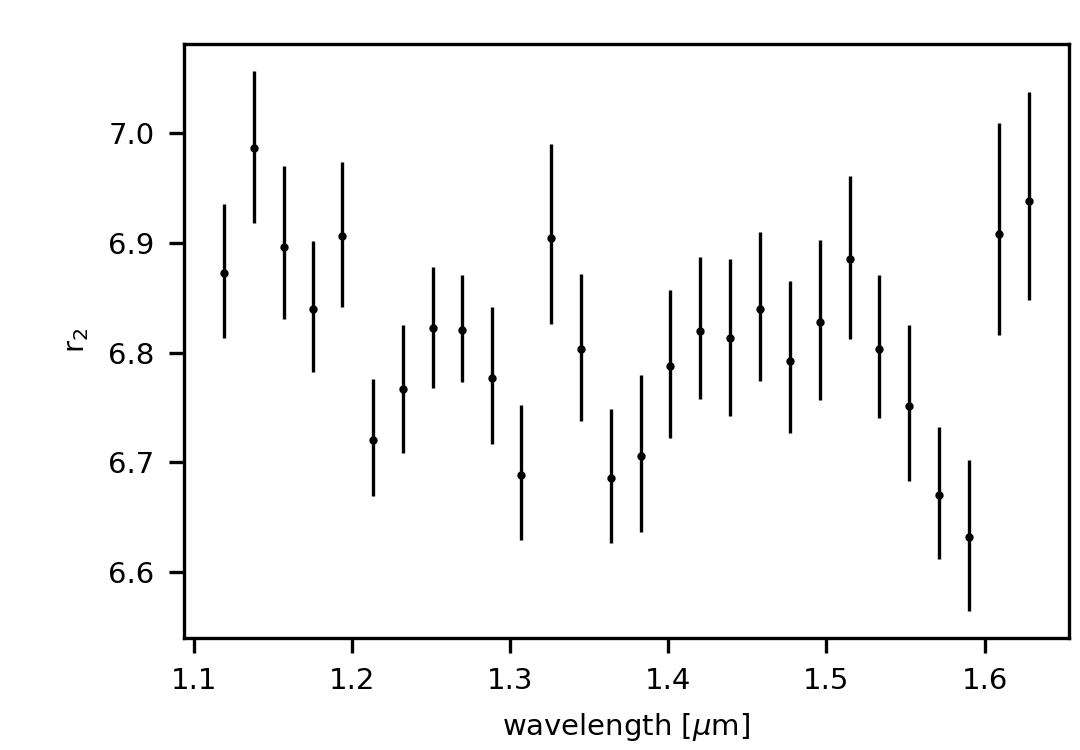}
         \label{fig:r2}
     \end{subfigure}
\vspace*{-10mm}
\caption{Variation of the fitted "model\_ramp" parameters $r_1$ and $r_2$ (cf. \cref{eq:ramp}) with wavelength. The dots represent best fit values in each wavelength bin with $\SI{1}{\sigma}$ errors from the posterior. Both $r_1$ and $r_2$ are clearly not wavelength-independent.}
\label{fig:model_ramp_parameters_wvl}
\end{figure*}

We plot the transmission spectrum as the transit depth versus the wavelength of the spectroscopic light curve bins. For the "model\_ramp" fits to the spectroscopic light curves, we found that the exponential ramp parameters and the linear slope varied with wavelength (see \cref{fig:model_ramp_parameters_wvl}). Since the "divide\_white" method relies on the assumption that the systematics are independent of wavelength, we used the "model\_ramp" spectrum for the following retrieval analyses. The transmission spectra obtained with both models for the instrument systematics are shown in \cref{fig:spec}. This figure also illustrates that the assumption of wavelength-independent systematics ("divide\_white" spectrum) leads to a slope in the spectrum, which could potentially bias atmospheric retrievals.

\begin{figure}[!htb]
\centering
\includegraphics[scale=1]{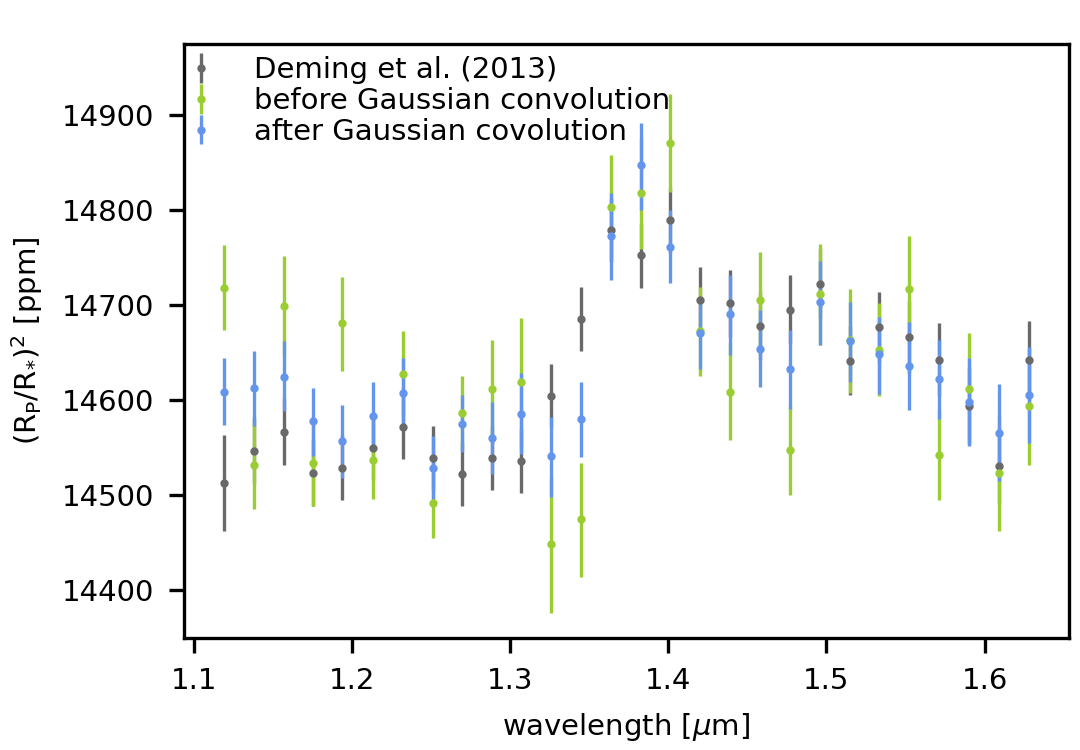}
\vspace{-6mm}
\caption{Comparison of HST/WFC3 transmission spectra with (blue) and without (green) applying the additional Gaussian convolution (see \cref{subsec:Deming_method}) in the data reduction. Both spectra use the "model\_ramp" model. The green spectrum without the additional smoothing step shows a "zig-zag" pattern of alternating transit depths between subsequent exposures. The grey spectrum is from \citet{Deming_2013} for comparison.}
\label{fig:zigzag}
\vspace*{-4mm}
\end{figure}

\subsection{Recreation of the \citet{Deming_2013} data reduction method}
\label{subsec:Deming_method}
In the original reduction, we noticed a "zig-zag" pattern (see \cref{fig:zigzag}) of alternating higher and lower transit depths between subsequent exposures, particularly at shorter wavelengths. This pattern was not present in the original transmission spectrum by \citet{Deming_2013}. To investigate the origin of this pattern, we recreated the reduction method used by \citet{Deming_2013}. As they were the first to report the extraction of an atmospheric transmission spectrum from spatially scanned transit observations of an exoplanet, they employed a slightly different method than the one commonly used today. The main differences between the two methods are explained below.

After extracting a 1D spectrum at each exposure, as in the current method, \citet{Deming_2013} convolved each of these spectra with a Gaussian kernel of FWHM = $\SI{4}{\pix}$ to reduce the effect of undersampling the spectrum and avoid degradation of the spectral resolution. The template spectrum for the wavelength correction was not derived from a modelled stellar spectrum but through averaging the spectra one hour before the start of the transit and one hour after the end. Furthermore, all individual spectra were resampled to a finer wavelength sampling and then interpolated to a common wavelength solution. At each exposure, the template spectrum is fit separately to the individual spectrum via least-squares fitting by shifting it in wavelength and stretching it in flux. The residuals of this fit, i.e. fit minus template spectrum, were binned into the same 28 wavelength channels as mentioned earlier. We then fit an eclipse model from the \texttt{batman} code \citep{Kreidberg_2015} and a linear baseline to these residuals. Here, the eclipse model is used instead of the transit model to fit declines and inclines of the flux in the residuals. The transmission spectrum is generated by adding the eclipse depth at each binned wavelength to the white light transit depth. We were successful in recreating this method.

\begin{figure*}
\centering
\includegraphics[scale=1]{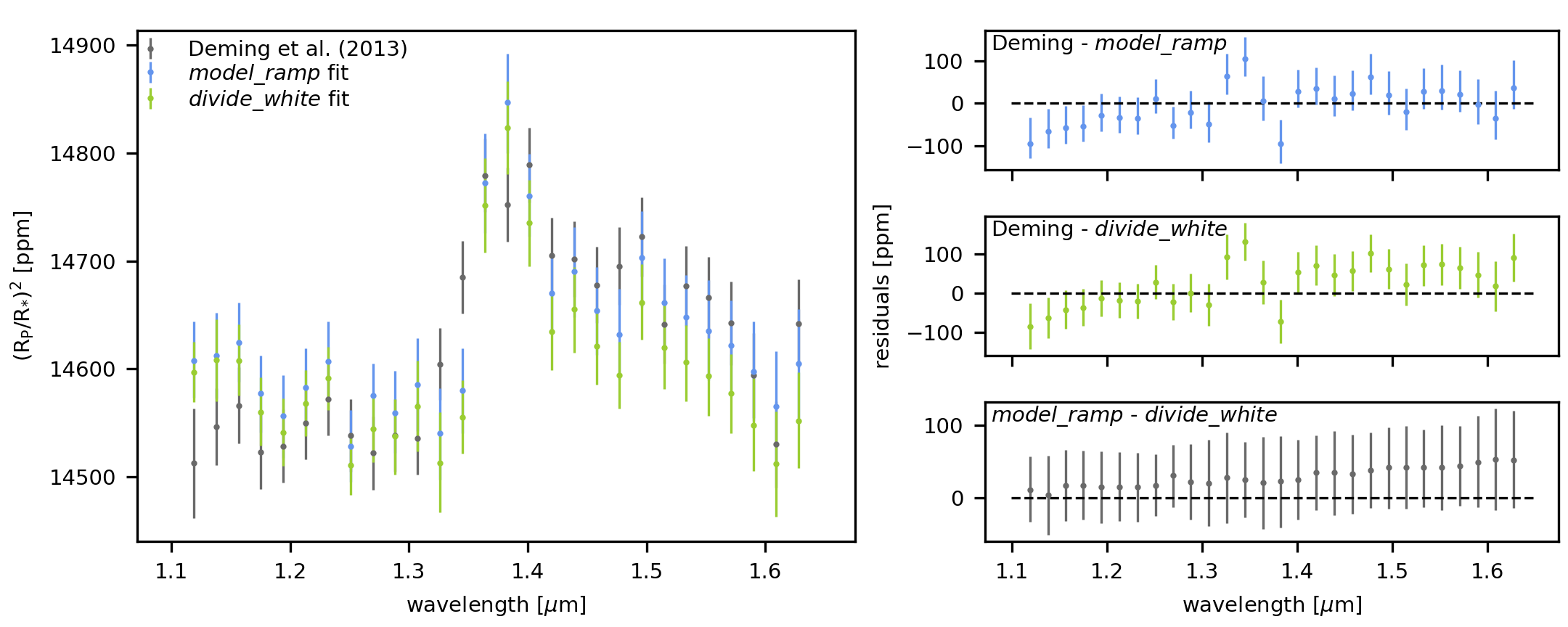}
\caption{Comparison of HST/WFC3 transmission spectra for three different data reduction and fitting methods for \planet. The blue spectrum uses exponential functions to fit the systematics at the beginning of each orbit ("model\_ramp" fit), whereas the green one assumes wavelength independent systematics ("divide\_white" fit). The grey spectrum is the original transmission spectrum published by \citet{Deming_2013}. The right panels show the differences between all three spectra. Noticeable is the linear trend with wavelength in the residuals for all three cases.}
\label{fig:spec}
\vspace*{-3mm}
\end{figure*}

While working on this method, we realised that smoothing the individual spectra at each exposure with a Gaussian kernel was the crucial step to avoid the "zig-zag" pattern, which also appeared in the transmission spectrum generated with the reduction method of \citet{Deming_2013} when the smoothing was turned off. As a result, we implemented this step in the \texttt{PACMAN} code. \citet{Deming_2013} included this additional smoothing originally to reduce undersampling effects due to changes in the stellar line shapes during the transit. These could also be the reason for the pattern we see in our transmission spectra before adding the Gaussian convolution. However, \citet{Deming_2013} noted most pronounced differences near strong stellar lines, like Paschen-$\beta$ ($\lambda$ = \SI{1.28}{\micro\meter}) and we do not see strong effects there (see \cref{fig:zigzag}). Nevertheless, the added Gaussian convolution step helps to avoid the "zig-zag" pattern. Figure~\ref{fig:zigzag} shows a comparison of the "model\_ramp" spectra before and after applying the Gaussian convolution, and \cref{fig:spec} shows the transmission spectrum obtained using both models for the instrument systematics compared to the transmission spectrum of \citet{Deming_2013}.

%--------------------------------------------------------------------
\section{Atmospheric retrievals of the newly reduced HST spectrum}
\label{sec:HST_retrievals}
Atmospheric retrievals are a powerful tool for analysing transmission spectra. For this study, we used \texttt{petitRADTRANS} to perform atmospheric retrievals, which incorporates the \texttt{pymultinest} (\cite{Buchner_2014}) nested sampling algorithm. The chosen models and the parameter setup will be introduced in the following \cref{subsec:retrieval_setup}. \hyperref[subsec:retrieval_results]{Section~\ref{subsec:retrieval_results}} will summarise the results of the atmospheric retrieval analyses of the newly reduced HST/WFC3 transmission spectrum.

\subsection{Model and parameter setup}
\label{subsec:retrieval_setup}
Several previous studies, e.g. \citet{Madhusudhan_2014}, \citet{Benneke_2015}, \citet{Barstow_2017}, \citet{MacDonald_2017}, and \citet{Xue_2023}, have run retrievals on \planet's transmission spectrum. The results, especially the water abundances obtained, were contradictory, as mentioned earlier. In the following, we describe the model and parameter setup for atmospheric retrievals in this paper. As stated above, we used the "model\_ramp" transmission spectrum for the retrievals. For comparison, we also ran retrievals on the original spectrum published by \citet{Deming_2013}. Using the planetary radius $R_{\rm pl} = \SI{1.359}{R_{Jup}}$ and mass $M_{\rm pl} = \SI{0.682}{M_{Jup}}$ \citep{Bonomo_2017}, we retrieve a reference pressure $P_{\rm ref}$ at which this radius and mass are reached, together with the other model parameters. This method has the advantage of using "directly" measurable quantities for the planet instead of the inferred gravity log\,($g$). 

We assumed an isothermal atmosphere for the planet and tested cloud-free models, patchy clouds, and a completely opaque grey cloud deck. \cref{tab:ret_params} shows the parameter inputs adopted for the forward models, and \cref{tab:clouds} provides an overview of all the cloud models tested in this study. \cref{tab:ret_list} provides a list of all retrievals run on the spectra with the results. 

We used a free chemistry retrieval approach with $\mathrm{H_2O}$, $\mathrm{CH_4}$, $\mathrm{NH_3}$, $\mathrm{HCN}$, and $\mathrm{Na}$ as line-absorbing species in the atmosphere, given by the main absorbers in the wavelength range of the transmission spectrum and the claimed findings of previous work \citep{MacDonald_2017, Tsiaras_2018, Xue_2023}. We used a model resolution of 200, which is sufficient for the resolution of our data (less than R=100). For the rebinning of the opacities we used the \texttt{Exo\_k} package \citep{Leconte_2021} in \texttt{petitRADTRANS}' "correlated-k" mode. We used opacities calculated from the \textit{Pokazatel} \citep{POKAZATEL} line list for $\mathrm{H_2O}$, \textit{HITEMP} \citet{HITEMP} for $\mathrm{CH_4}$, the lists by \citet{Exomol_NH3} for $\mathrm{NH_3}$ and by \citet{Exomol_HCN} for $\mathrm{HCN}$, and the \citet{VALD} line list for $\mathrm{Na}$. For all line species, we adopted log-uniform priors with mass fraction abundance limits of $10^{-12}$ to $1$. Hydrogen and helium were assumed to be responsible for Rayleigh scattering in the atmosphere and also contribute to continuum opacities via collision-induced absorption of $\mathrm{H_2-H_2}$ and $\mathrm{H_2-He}$. We assumed an atmosphere consisting mainly of hydrogen and helium with mass mixing ratios of $0.74$ and $0.24$, respectively. Due to the fairly small size of the retrievals, we used 1000 live points.

\begin{table}[!htb]
\centering
\caption[Input parameters for HST retrievals]{Input parameters for the retrievals of the HST spectra.}
\begin{tabular}{c l}
 \hline
 \multicolumn{1}{c}{parameter} & \multicolumn{1}{c}{prior} \\
 \hline \hline
 log\,($P_{\rm ref}$) [\SI{}{log\,(\bar)}] & $U\,(-6.0, 3.0)$ \\
 $T \: [\SI{}{\kelvin]}$ & $U\,(800, 1300)$ \\
 mass fraction log\,($X$) & $U\,(-12.0, 0.0)$ \\
 log\,($P_{\rm cloud}$) [\SI{}{log\,(\bar)}] & $U\,(-6.0, 3.0)$\\
 cloud fraction & $U\,(0.0, 1.0)$\\
 \hline
\end{tabular}
\tablefoot{For the radius of the host star, we adopted the value $R_* = \SI{1.155}{R_\odot}$ \citep{Bonomo_2017}. The planet's radius $R_{\rm pl} = \SI{1.359} {R_{Jup}}$ and mass $M_{\rm pl} = \SI{0.682}{M_{Jup}}$ were fixed and taken from the same source. For the planet's reference radius, we used a Jupiter radius of $R_{\rm Jup} = \SI{6.9911e7}{\metre}$. All retrieved parameters have uniform/log-uniform priors ($U$). Additionally, different absorbing species were included in the retrievals (see \cref{tab:ret_list}).}
\label{tab:ret_params}
\vspace*{-2mm}
\end{table}

\begin{table}[!htb]
\centering
\caption{Overview for the different cloud models used in this study.}
\centering
\begin{tabular}{>{\raggedright\arraybackslash}p{1.95cm}|>{\raggedright\arraybackslash}p{6.3cm}}
 \hline
 \multicolumn{1}{l}{cloud model} & \multicolumn{1}{c}{description} \\
 \hline \hline
 no clouds & clear atmosphere without any cloud opacity \\
 clouds & fully opaque atmosphere at pressures $P > P_{\rm cloud}$ \\
 patchy clouds & similar to clouds but only covering a fraction of the atmosphere \citep[][]{Line_Parmentier_2016} \\
 complex clouds & cloud deck with opacity given by \cref{eq:cloud_opac} and cloud base at pressure $P_{\rm base}$ \\
 complex patchy clouds & similar to complex clouds but only covering a fraction of the atmosphere \\
 \hline
\end{tabular}
\tablefoot{The input priors for the different parameters are listed in \cref{tab:ret_params,tab:joint_ret_params,tab:params_cc}.}
\label{tab:clouds}
\vspace*{-3mm}
\end{table}

\subsection{Retrieval results}
\label{subsec:retrieval_results}
\begin{figure}[!htb]
\centering
\includegraphics[scale=1]{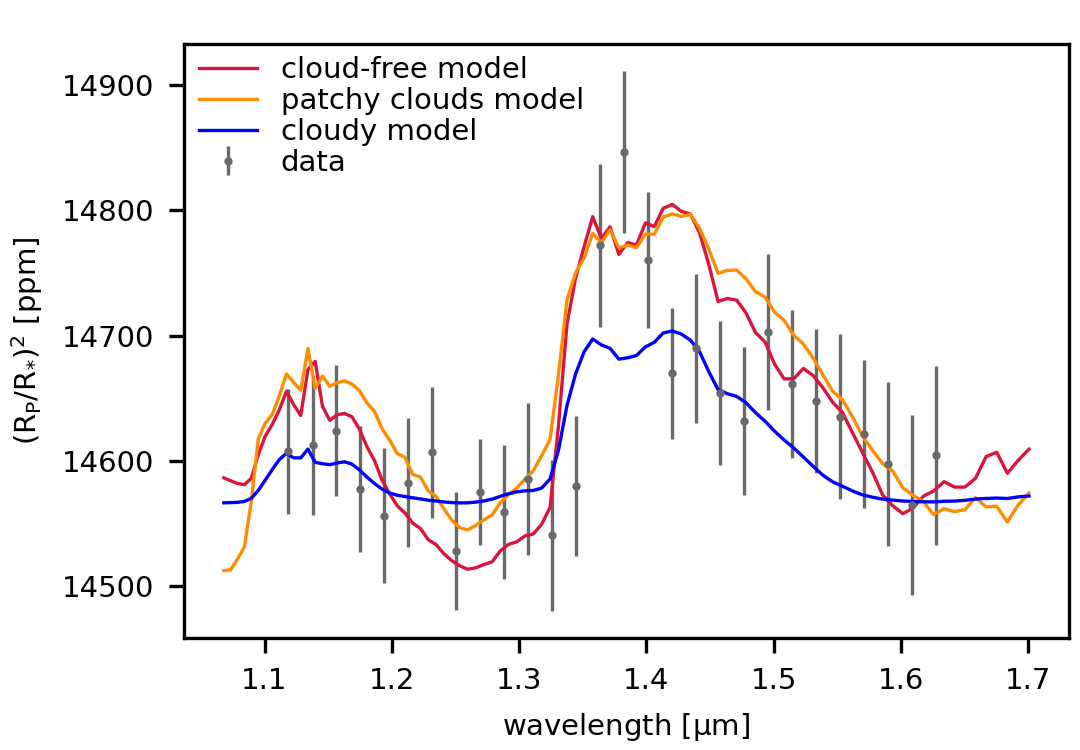}
\vspace{-6mm}
\caption{Comparison of best-fit spectra with different cloud assumptions. The retrieval uses the "model\_ramp" fit HST/WFC3 spectrum (grey dots). All three retrievals include all tested absorbing species (see \cref{tab:ret_list}). The red spectrum is cloud-free, the blue one cloudy, and the orange spectrum is the best fit for patchy clouds.}
\label{fig:HST_mr_cloud_comparison}
\end{figure}

A list of all retrievals on the different HST spectra can be found in \cref{tab:ret_list}. To robustly estimate the various parameters and compare the different models, we used a Bayesian framework. The Bayesian evidence quantifies the suitability of the tested atmospheric model to the data \citep{MacDonald_2017}. The Bayes factor $Z_{\rm ref}/Z$ gives an indication of which model is favoured, and factors of 3, 12, and 150 are often interpreted as "weak", "moderate", and "strong" detections of the reference over the other model \citep{Trotta_2008, Benneke_2013}. Bayes factors can also be transformed into $\sigma$-significances, using Eqs. 8 and 9 in \citet{Benneke_2013}. With this approach, we first determined the best-fitting atmospheric model from cloudy, patchy, or no cloud models, using all line species in the retrievals. As \cref{tab:ret_list} shows, all three models have similar Bayesian evidence and Bayes factors, which are too small to even weakly distinguish one model from the others. The best-fit spectra for these models are shown in \cref{fig:HST_mr_cloud_comparison}. The clouds expected on a planet like \planet{} could consist of $\mathrm{Fe, Ni, Al_2O_3, Mg_2SiO_4, MgSiO_3, MgFeSiO_4, SiC, TiO_2}$ or $\mathrm{MnS}$ \citep{Burrows_1999, Benneke_2015, Mbarek_2016, Woitke_2018, Kitzmann_2024} assuming a solar-composition gas.

\begin{figure}[!htb]
\centering
\includegraphics[scale=1]{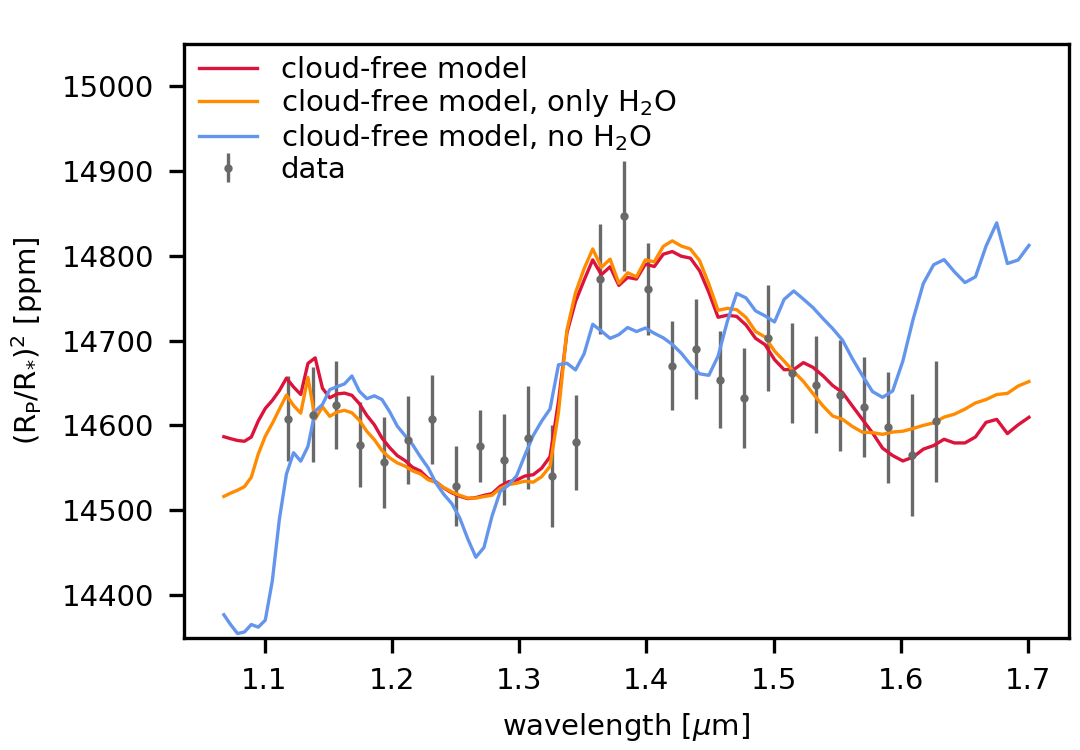}
\vspace{-6mm}
\caption{Comparison of best-fit cloud-free spectra for retrievals of the "model\_ramp" fit HST/WFC3 spectrum. The red spectrum is the same as in \cref{fig:HST_mr_cloud_comparison}, the orange spectrum is the best fit for the model with the largest Bayesian evidence (clear atmosphere and only $\mathrm{H_2O}$ as line absorber). The light blue spectrum is the model without containing water as an absorber.}
\label{fig:HST_mr_H2O_comparison}
\end{figure}

In the further analysis of this spectrum, we used the cloud-free model as a reference because it has fewer parameters than the cloudy or patchy clouds model. As the Bayes factors do not clearly distinguish between them, we prefer the simplest possible model, which also has the lowest $\chi^2_{red}$ value. The different line-absorbing species were tested by running retrievals in which one absorber was left out to determine how well it was detected. As \cref{tab:ret_list} shows, H$_2$O was the only species detected, with a significance of $\SI{4}{\sigma}$ (calculated from Bayes factors). None of the other absorbers are significantly detected. Moreover, we compared the full model with all absorbing species to one that included only water; we found that the latter was moderately favoured by a Bayes factor of 5, further emphasising the fact that water is the only relevant line absorber in this wavelength range. Best-fit spectra for several of the tested models are shown in \cref{fig:HST_mr_H2O_comparison}. $\mathrm{H_2O}$ is clearly recognisable as the main absorbing molecular species in this wavelength range. 

To combine the results for the parameters of all different retrieval models, we applied the Bayesian model averaging (BMA) method \citep{Nixon_2024, Nasedkin_2024_BMA}. For parameters that are shared between several models, we obtain a combined posterior distribution. For the calculation, the posterior distributions of different models are summed with weights according to their Bayesian evidence \citep[see Eqs. 12 - 14 in][]{Nasedkin_2024_BMA}. This results in more robust uncertainties for every parameter as the uncertainties from the data and prior distribution as well as the model uncertainty itself are taken into account. To avoid bias in the results, all the models used in the averaging have to be conceptually different. Therefore, we average only models with different cloud treatments, including all species, since most line absorbers do not have constrained abundances. Averaging over all models would possible bias the results towards the cloud-free case, as these make up the majority of the models. In \cref{fig:BMA_HST}, we show the averaged posterior distributions for the parameters present in most models. Since the abundances of $\mathrm{CH_4}$, $\mathrm{NH_3}$, $\mathrm{HCN}$, and $\mathrm{Na}$ are not constrained, we only report $\SI{3}{\sigma}$ upper limits of $\mathrm{log(\chi_{CH_4})} = -3.43$, $\mathrm{log(\chi_{NH_3})} = -2.62$, $\mathrm{log(\chi_{HCN})} = -2.66$, and $\mathrm{log(\chi_{Na})} = -2.01$. We could not reproduce former detections of methane or nitrogenous species, as claimed in \citet{MacDonald_2017}. The retrieved values for all parameters can be found in the \hyperref[sec:appendix]{appendix}, \cref{tab:ret_complete_results}.

For comparison, we ran the same retrievals using the original transmission spectrum published by \citet{Deming_2013}. The retrievals show similar results for the $\mathrm{H_2O}$ abundance, but the $\chi^2_{\rm red}$ values of the retrievals using our data reduction are much smaller (1.33 vs 1.99 for the preferred cloud-free, only $\mathrm{H_2O}$ model, cf. \cref{tab:ret_list}). This shows that the new data reduction indeed has an effect and might change the results, as we obtain better model fits to our transmission spectrum.

\renewcommand{\arraystretch}{1.5}
\begin{table*}
\caption{Summary of all retrievals run on the HST spectra.}
\centering          
\begin{tabular}{c c c c c c c}
\hline
spectrum & retrieval model & line species & \parbox[c][25pt]{2.7cm}{water abundance \\ log\,($\chi_{\rm{H_2O}}$) in VMR} & $\chi^2_{\rm red}$ & \parbox[c][25pt]{2cm}{Bayesian evi-\\dence log\,($Z$)} & $Z_{\rm ref}/Z$\\
 \hline \hline
 "model\_ramp" & clouds & all & $-4.33^{+1.86}_{-1.16}$ & 1.66 & $101.1\pm0.2$ & $3.16$ \\
 "model\_ramp" & patchy clouds & all & $-4.97^{+2.06}_{-0.55}$ & 1.95 & $100.0\pm0.1$ & $3.98$ \\
 "model\_ramp" & no clouds & all & $-5.50^{+0.26}_{-0.27}$ & 1.49 & $\hphantom{0} 99.9\pm0.1$ & $5.01$ \\
 "model\_ramp" & no clouds & only $\mathrm{H_2O}$ & $-5.50^{+0.42}_{-0.27}$ & 1.33 & $100.6\pm0.0$ & ref \\
 "model\_ramp" & clouds & only $\mathrm{H_2O}$ & $-5.12^{+2.95}_{-0.53}$ & 1.37 & $100.5\pm0.0$ & 1.26 \\
 "model\_ramp" & no clouds & no $\mathrm{H_2O}$ & \ldots & 2.21 & $\hphantom{0} 96.5\pm0.1$ & $12589.3$ \\
 "model\_ramp" & clouds & no $\mathrm{H_2O}$ & \ldots & 1.98 & $\hphantom{0} 96.8\pm0.1$ & $6309.6$ \\
 "model\_ramp" & no clouds & no $\mathrm{CH_4}$ & $-5.41^{+0.36}_{-0.27}$ & 1.48 & $100.2\pm0.1$ & $2.51$ \\
 "model\_ramp" & no clouds & no $\mathrm{NH_3}$ & $-5.51^{+0.27}_{-0.27}$ & 1.67 & $100.0\pm0.1$ & $3.98$ \\
 "model\_ramp" & no clouds & no $\mathrm{HCN}$ & $-5.51^{+0.26}_{-0.28}$ & 1.51 & $\hphantom{0} 99.9\pm0.0$ & $5.01$ \\
 "model\_ramp" & no clouds & no $\mathrm{Na}$ & $-5.49^{+0.28}_{-0.27}$ & 1.57 & $\hphantom{0} 99.9\pm0.1$ & $5.01$ \\
\hline
 \citet{Deming_2013} & clouds & all & $-5.05^{+1.62}_{-0.37}$ & 2.46 & $101.6\pm0.0$ & $3.16$ \\
 \citet{Deming_2013} & patchy clouds & all & $-5.12^{+0.51}_{-0.30}$ & 2.51 & $102.0\pm0.2$ & $1.26$ \\
 \citet{Deming_2013} & no clouds & all & $-5.24^{+0.26}_{-0.28}$ & 2.22 & $101.9\pm0.1$ & $1.58$\\
 \citet{Deming_2013} & no clouds & only $\mathrm{H_2O}$ & $-5.58^{+0.15}_{-0.13}$ & 1.99 & $102.1\pm0.1$ & ref\\
 \citet{Deming_2013} & no clouds & no $\mathrm{H_2O}$ & \ldots & 4.36 & $\hphantom{0}91.0\pm0.0$ & $1.26\cdot10^{11}$\\
 \citet{Deming_2013} & clouds & only $\mathrm{H_2O}$ & $-5.57^{+0.16}_{-0.13}$ & 2.08 & $101.6\pm0.1$ & 3.16\\
 \hline
\label{tab:ret_list}
\end{tabular}
\vspace*{-7mm}
\tablefoot{The model with the largest Bayesian evidence is the one labelled no clouds, only $\mathrm{H_2O}$. For a description of the cloud models see \cref{tab:clouds}. Results for all parameters are summarised in \cref{tab:ret_complete_results}.}
\vspace*{-5mm}
\end{table*}

\begin{figure}[!htb]
\centering
\includegraphics[scale=1]{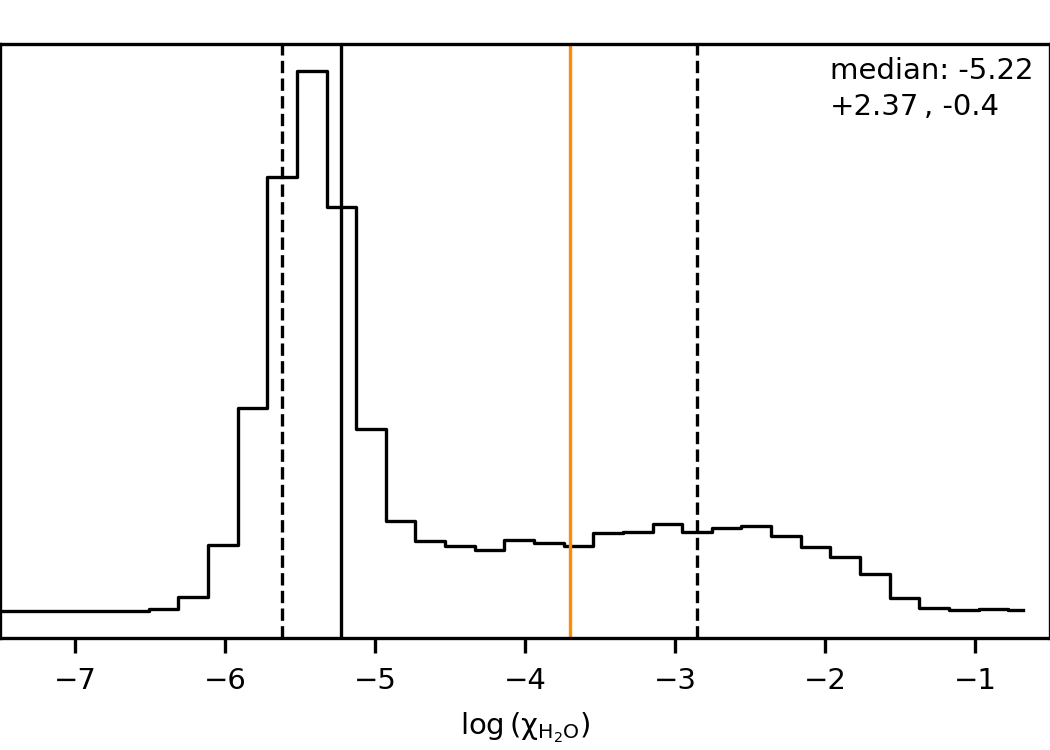}
\vspace*{-2mm}
\caption{Bayesian model averaged posterior distribution for $\mathrm{H_2O}$ from the HST/WFC3 retrievals. The black line indicates the median of the distribution, and the dashed lines are the $\pm34.1\%$ confidence regions. The orange lines shows the solar value ($\mathrm{log(\chi_{H_2O})}$ = -3.70) at $\SI{1}{\milli\bar}$ and $\SI{1200}{\kelvin}$. Since most of the models used for the BMA do not include clouds, the results are more heavily weighted toward the cloud-free solution and therefore lower $\mathrm{H_2O}$ abundances (see text).}
\label{fig:BMA_HST_H2O}
\end{figure}

Figure~\ref{fig:BMA_HST_H2O} shows that we retrieved a Bayesian averaged water abundance of $\mathrm{log(\chi_{H_2O})} = -5.22^{+2.37}_{-0.40}$ from the HST spectrum. This is well below solar abundances and is consistent with several previous studies \citep{Madhusudhan_2014, Barstow_2017, MacDonald_2017}. The low water abundance can be explained by the use of cloud-free models when retrieving the HST spectrum. Since the water feature in the spectrum is quite small and in most models no clouds are present to suppress it, a lower water abundance is needed to fit the data than when a cloud is present \citep{Welbanks_2021}. This trend can be seen in \cref{tab:ret_list} as well as in \cref{tab:joint_ret_results} for the joint HST-JWST retrievals. As \citet{Madhusudhan_2014_b} state, such a sub-solar water abundance could be explained by a significant sub-solar metallicity; alternatively a highly super-solar C/O would be required. From the JWST and the joint HST-JWST spectrum (see \cref{subsec:joint_ret_results}) as well as other hot Jupiter data sets \citep[e.g.][]{Ahrer_2022, Fu_2024, Bell_2024}, we expect a cloudy rather than a clear atmosphere for \planet. Therefore, we also included retrievals using the cloudy model and just $\mathrm{H_2O}$ as a line absorber (\cref{tab:ret_list}). The water abundance matches those of the other retrievals and is  between the cloud-free and cloudy results when we used all species in the retrievals (see \cref{tab:ret_list}). Even though the cloudy models are not favoured by the Bayesian evidence or the $\chi^2_{\rm red}$ values from the HST data alone, they probably are the most realistic ones for \planet's atmosphere. They also show a clear tendency (cf. \cref{tab:ret_list}) to higher $\mathrm{H_2O}$ abundances, which we would expect. Since earlier analyses of the HST data set mostly favoured lower $\mathrm{H_2O}$ abundances, similar to our cloud-free ones and in contradiction to expectations, our study clearly shows the limitations of obtaining robust results when using only HST data due to the limited wavelength coverage and precision. Here, adding JWST data with better precision and broader wavelength coverage leads to more robust results (see also \cref{subsec:joint_ret_results}). In addition, we caution against using only Bayesian evidence as a criterion to decide on most likely models, as it may favour simpler but not necessarily more realistic models. Planetary atmospheres are inherently complex and might not be best described by simpler models, even if statistically preferred.

%--------------------------------------------------------------------
\section{Atmospheric retrievals of the combined HST-JWST spectrum}
\label{sec:comb_retrievals}

As shown in \cref{fig:spec}, the new data reduction used in this study yields slightly different results compared to the original spectrum, which makes it worthwhile to conduct a joint atmospheric analysis of this planet again. The broader wavelength coverage of the combined data set also allows us to test a broader range of retrieval approaches, including different and more complex cloud models, and adding species that do not absorb in the HST wavelength range. 

\subsection{Joint retrieval setup}
\label{subsec:joint_ret_setup}
JWST/NIRCam operates in a wavelength range different from HST's WFC3, so we included $\mathrm{CO_2}$ and $\mathrm{CO}$ in the list of line absorbers for the joint retrieval. Because several years passed between the observations, differences in stellar activity cannot be excluded. The two instruments are also affected by different systematics. To account for these differences, we also introduced an optional offset with a Gaussian prior (which covers the offset retrieved by \citep{Xue_2023}) for the HST data relative to the JWST data in the parameters. For the temperature, we adopted a broader uniform prior and included the planet's mass and radius \citep{Bonomo_2017} along with the associated errors as retrieved parameters in the setup. These additional free parameters were not feasible for retrieving the HST data alone due to the limited number of data points. The results we obtain are not influenced by the choice of narrow uniform priors for the planet's mass and radius. We got comparable results for all parameters when using wider Gaussian priors, with the exception of the reference pressure as expected. Since the retrievals using the narrow priors have a shorter runtime, we continued using them. For the atmospheric model, we used the same isothermal transmission model as described in \cref{subsec:retrieval_setup}, with the same line lists for the absorbers and the \textit{HITEMP} \citep{HITEMP} list for $\mathrm{CO}$ and the list by \citet{Exomol_CO2} for $\mathrm{CO_2}$. The input parameters for the joint retrievals are provided in \cref{tab:joint_ret_params}. Due to the larger data set, we used 4000 live points for the joint retrievals. We ran the retrievals on the "model\_ramp" HST data, combined with the JWST/NIRCam data published by \citet{Xue_2023}. 

\begin{table}[h]
\centering
\caption{Input parameters for the joint retrievals on the combined HST-JWST spectra.}
\begin{tabular}{c l}
 \hline
 \multicolumn{1}{c}{parameter} & \multicolumn{1}{c}{prior} \\
 \hline \hline
 $R_{\rm pl} \: [\SI{}{R_{Jup}}]$ & $U\,(1.340, 1.375)$ \\
 $M_{\rm pl} \: [\SI{}{M_{Jup}}]$ & $U\,(0.667, 0.696)$ \\
 log\,($P_{\rm ref}$) [\SI{}{log\,(\bar)}] & $U\,(-6.0, 3.0)$ \\
 $T \: [\SI{}{\kelvin}]$ & $U\,(500, 1600)$ \\
 mass fraction log\,($X$) & $U\,(-12.0, 0.0)$ \\
 log\,($P_{\rm cloud}$) [\SI{}{log\,(\bar)}] & $U\,(-6.0, 3.0)$ \\
 cloud fraction & $U\,(0.0, 1.0)$ \\
 offset & $N\,(0.0, 10^{-3})$ \\
 \hline
\end{tabular}
\tablefoot{For the radius of the host star, we used the value $R_* = \SI{1.155}{R_\odot}$ \citep{Bonomo_2017}. The priors for the planets radius $R_{\rm pl}$ and mass $M_{\rm pl}$ we took from the same source. The planet's reference radius is given in Jupiter radii ($R_{\rm Jup} = \SI{6.9911e7}{\metre}$). All retrieved parameters have uniform/log-uniform priors ($U$), except the offset, which has a Gaussian prior ($N$). Additionally, we included different absorbing species in the retrievals (see \cref{tab:joint_ret_results}).}
\label{tab:joint_ret_params}
\end{table}

\subsection{Joint retrieval results}
\label{subsec:joint_ret_results}
\cref{tab:joint_ret_results} presents the results of the combined retrievals. To determine the best model, we used the approach described in \cref{subsec:retrieval_results}. Since the cloudy model has a slightly larger Bayesian evidence log\,($Z$) than the patchy cloud model, we selected the cloudy one for further analysis, as it also has one parameter less, which is not strictly necessary to fit the data. Both models show a reasonably strong preference ($\SI{3.6}{\sigma}$) over a clear atmosphere. Best-fit spectra for all three models are shown in \cref{fig:comb_cloud_model_comparison}.

For the cloudy models, we evaluated which line species are required to fit the data properly. By comparing the Bayesian evidence, only $\mathrm{H_2O}$ and $\mathrm{CO_2}$ can be robustly claimed to be detected (\cref{tab:joint_ret_results}). The cloudy model with only these two molecules as line absorbers has the largest Bayesian evidence of all tested models. This further supports the conclusion that all other absorbing species are unnecessary for interpreting the data. Water is detected at over $\SI{14}{\sigma}$ (for the corresponding Bayes factors, see \cref{tab:joint_ret_results}) with a log volume mixing ratio of $\mathrm{log(\chi_{H_2O})} = -4.47^{+1.26}_{-0.50}$ in the cloudy model and $\mathrm{log(\chi_{H_2O})} = -4.51^{+1.15}_{-0.48}$ in the model that only includes $\mathrm{H_2O}$ and $\mathrm{CO_2}$. $\mathrm{CO_2}$ is detected at more than $\SI{7}{\sigma}$ (see \cref{tab:joint_ret_results}) with a log volume mixing ratio of $\mathrm{log(\chi_{CO_2})} = -7.64^{+1.18}_{-0.37}$ in the cloudy model and $\mathrm{log(\chi_{CO_2})} = -7.67^{+1.07}_{-0.36}$ in the model that only includes $\mathrm{H_2O}$ and $\mathrm{CO_2}$. The model with the highest evidence (clouds and only $\mathrm{H_2O}$ and $\mathrm{CO_2}$ as line absorbers) is preferred over the one that includes all other species tested by $\SI{3.1}{\sigma}$. All other tested species are not required to fit the spectrum with significances ranging from $\SI{2.8}{\sigma}$ to $\SI{3.0}{\sigma}$, and they do not show constrained posterior distributions. Therefore, we only report upper limits for the volume mixing ratios of all these species (see \cref{subsec:BMA_joint_ret}). Figure~\ref{fig:comb_species_comparison} shows the best-fit spectra of some of the tested models, and \cref{fig:comb_corner_plot_best_ret} presents the corner plot for the model with the largest Bayesian evidence, the cloudy model including only $\mathrm{H_2O}$ and $\mathrm{CO_2}$. 

\citet{Tsai_2023} detected photochemically-produced $\mathrm{SO_2}$ in the atmosphere of the hot Jupiter WASP-39\,b. Since this molecule could also be present in the atmospheres of other hot Jupiters, we included it as an absorber in the analysis, using the line list from \citet{Exomol_SO2}. We only retrieved a $\SI{3}{\sigma}$ upper limit of $\mathrm{log(\chi_{SO_2})} = -5.93$ for the volume mixing ratio. The Bayes factor also disfavours the inclusion of $\mathrm{SO_2}$ in the species required to describe the atmosphere of \planet{} with a significance of $\SI{1.82}{\sigma}$ compared to the cloudy model and $\SI{3.3}{\sigma}$ compared to the cloudy model that only includes $\mathrm{H_2O}$ and $\mathrm{CO_2}$, which has the largest Bayesian evidence.

We also tested $\mathrm{H_2S}$, another sulphur-bearing species expected at the planetary temperature of \planet. \citet{Xue_2023} reported a tentative hint of its presence in the atmosphere, and $\mathrm{H_2S}$ was previously detected on the Hot Jupiter HD 189733\,b \citep{Fu_2024}. Our retrievals using the ExoMol line list \citep{Exomol_H2S} do not yield a constrained posterior distribution for $\mathrm{H_2S}$. We obtained a $\SI{3}{\sigma}$ upper limit for its volume mixing ratio of $\mathrm{log(\chi_{H_2S})} = -4.00$, with Bayesian evidence favouring the cloudy model without it (see \cref{tab:joint_ret_results}).

The measured water abundances from the joint spectra are higher than those from the retrievals of the HST spectrum alone. Since the spectral range of the combined HST-JWST spectrum is wider than previously analysed (see \cref{sec:HST_retrievals}) and contains an additional water feature around $\SI{2.8}{\micro\meter}$ (cf. \cref{fig:opacity_contribution}), the water abundance is likely better constrainable and more reliable than the result obtained using only HST data. The results for all parameters and retrievals can be found in \cref{tab:joint_ret_complete_results} and \cref{tab:joint_ret_complete_results_2}.

\renewcommand{\arraystretch}{1.5}
\begin{table*}
\caption{Summary of all retrievals run on the combined HST-JWST spectra.}
\centering          
\begin{tabular}{c >{\centering\arraybackslash}p{2.3cm} c c c c c c}
\hline
retrieval model & line species & \parbox[c][25pt]{2.7cm}{water abundance \\ log\,($\chi_{\rm{H_2O}}$) in VMR} & \parbox[c][25pt]{1.5cm}{log\,($P_{\rm cloud}$) [log\,(bar)]} & \parbox[c][25pt]{1.5cm}{log\,($P_{\rm base}$) [log\,(bar)]} &  $\chi^2_{\rm red}$ & \parbox[c][25pt]{2cm}{Bayesian evi-\\dence log\,($Z$)} & $Z_{\rm ref}/Z$\\
 \hline \hline
 no clouds & all & $-4.39^{+0.24}_{-0.22}$ & \ldots & \ldots & 1.22 & $364.2\pm0.0$ & \ldots \\
 patchy clouds & all & $-4.25^{+0.90}_{-0.60}$ & $-2.16^{+0.68}_{-0.78}$ & \ldots & 1.02 & $366.1\pm0.0$ & \ldots \\
 clouds & all & $-4.47^{+1.26}_{-0.50}$ & $-1.45^{+0.27}_{-1.12}$ & \ldots & 1.06 & $366.3\pm0.0$ & $31.6$\\
 \hline
clouds & no $\mathrm{H_2O}$ & \ldots & $-0.65^{+0.08}_{-0.06}$ & \ldots & 3.36 & $322.1\pm0.0$ & $5.0\cdot10^{45}$\\
clouds & no $\mathrm{CH_4}$ & $-4.26^{+1.20}_{-0.60}$ & $-1.62^{+0.37}_{-1.09}$ & \ldots & 1.06 & $366.5\pm0.0$ & $20.0$\\
clouds & no $\mathrm{NH_3}$ & $-4.51^{+1.26}_{-0.47}$ & $-1.43^{+0.25}_{-1.11}$ & \ldots & 1.01 & $366.7\pm0.0$ & $12.6$\\
clouds & no $\mathrm{HCN}$ & $-4.52^{+1.23}_{-0.47}$ & $-1.44^{+0.25}_{-1.10}$ & \ldots & 1.03 & $366.7\pm0.0$ & $12.6$\\
clouds & no $\mathrm{Na}$ & $-4.46^{+1.29}_{-0.52}$ & $-1.47^{+0.28}_{-1.15}$ & \ldots & 1.02 & $366.6\pm0.2$ & $15.8$\\
clouds & no $\mathrm{CO_2}$ & $-2.58^{+1.70}_{-1.02}$ & $-2.19^{+1.92}_{-0.77}$ & \ldots & 1.74 & $354.3\pm0.0$ & $3.16\cdot10^{13}$\\
clouds & no $\mathrm{CO}$ & $-4.56^{+1.25}_{-0.46}$ & $-1.41^{+0.24}_{-1.09}$ & \ldots & 1.08 & $366.4\pm0.0$ & $25.1$\\
clouds & only $\mathrm{H_2O}$, $\mathrm{CO_2}$ & $-4.51^{+1.15}_{-0.48}$ & $-1.47^{+0.25}_{-1.00}$ & \ldots & 0.98 & $367.8\pm0.1$ & ref\\
clouds & with $\mathrm{SO_2}$ & $-4.35^{+1.27}_{-0.59}$ & $-1.53^{+0.34}_{-1.17}$ & \ldots & 1.04 & $366.0\pm0.0$ & $63.1$\\
clouds & with $\mathrm{H_2S}$ & $-4.37^{+1.30}_{-0.57}$ & $-1.52^{+0.33}_{-1.19}$ & \ldots & 1.09 & $366.0\pm0.0$ & $63.1$\\
clouds & more informed prior for $\mathrm{CO}$ & $-2.80^{+0.42}_{-0.41}$ & $-2.84^{+0.40}_{-0.40}$ & \ldots & 1.03 & $365.4\pm0.2$ & $251.2$ \\
\hline
complex clouds & all & $-2.81^{+0.47}_{-0.71}$ & \ldots & $-1.27^{+0.99}_{-0.78}$ & 1.02 & $365.9\pm0.0$ & \ldots \\
complex patchy clouds & all & $-3.48^{+0.71}_{-0.73}$ & \ldots & $-1.35^{+0.98}_{-0.85}$ & 1.05 & $366.1\pm0.0$ & \ldots \\
complex clouds & only $\mathrm{H_2O}$, $\mathrm{CO_2}$ & $-2.86^{+0.51}_{-0.96}$ & \ldots & $-1.17^{+0.95}_{-0.77}$ & 0.97 & $366.4\pm0.0$ & $25.1$ \\
\hline
\label{tab:joint_ret_results}
\end{tabular}
\vspace*{-7mm}
\tablefoot{The more complex cloud model is described by \cref{eq:cloud_opac}, and for this model. For a description of all cloud models see \cref{tab:clouds}. The model with the largest Bayesian evidence is the cloudy, only $\mathrm{H_2O}$ and $\mathrm{CO_2}$ one. Results for all the parameters and models are summarised in \cref{tab:joint_ret_complete_results} and \cref{tab:joint_ret_complete_results_2}.}
\end{table*}

\subsection{Retrievals with more complex clouds}
\label{subsec:complex_clouds}
Since the cloudy and patchy cloud models have almost identical Bayesian evidence, and the assumptions for the clouds are quite simple -- using a grey cloud deck at a retrieved pressure $P_{\rm cloud}$ in the atmosphere -- we also tested a more complex model for the clouds. We used the model described in \citet[Sect. 3.2 of the supplementary materials]{Dyrek_2023}. Their model employs the following description for the cloud opacity $\kappa_{\rm cloud}$:
\begin{equation}
\begin{aligned}
    \kappa_{\rm cloud} &= \frac{\kappa_{\rm base}}{1 + \left( \frac{\lambda}{\lambda_0} \right) ^p} \left( \frac{P}{P_{\rm base}} \right) ^{f_{\rm sed}} & \text{if} \: P < P_{\rm base} \ {\rm and }\label{eq:cloud_opac} \\
    \kappa_{\rm cloud} &= 0 & \text{if} \: P \geq P_{\rm base} {\rm .}
\end{aligned}
\end{equation}
Here, $\lambda$ represents the wavelength in micrometres and $P$ the pressure in bar. The free parameters also include the opacity at the cloud base $\kappa_{\rm base}$, the reference wavelength $\lambda_0$, the pressure at the cloud base $P_{\rm base}$, the power law coefficient $p$ (which describes the opacity dependence on the wavelength at $\lambda \gg \lambda_0$), and the sedimentation efficiency of the cloud particles $f_{\rm sed}$. The input values for these cloud parameters are listed in \cref{tab:params_cc}. For all other retrieved parameters, we used the same priors as before (cf. \cref{tab:joint_ret_params}), with the exception of the cloud pressure log\,($P_{\rm cloud}$), which was excluded. 

\begin{table}[H]
\centering
\caption{Input parameters for the more complex cloud model described in \cref{eq:cloud_opac} and used in joint retrievals on the combined HST-JWST spectra.}
\begin{tabular}{c l}
 \hline
 \multicolumn{1}{c}{parameter} & \multicolumn{1}{c}{prior} \\
 \hline \hline
 log\,($\lambda_0$) [\SI{}{log \,(\micro\meter)}] & $U\,(-4.0, 2.0)$ \\
 log\,($\kappa_{\rm base}$) $\left[\SI{}{log \left(\frac{\cm^2}{\gram}\right)}\right]$ & $U\,(-20, 20)$ \\
 log\,($P_{\rm base}$) [\SI{}{log\,(\bar)}] & $U\,(-6.0, 3.0)$ \\
 $p$ & $U\,(0, 6)$ \\
 $f_{\rm sed}$ & $U\,(0, 10)$ \\
 \hline
\end{tabular}
\tablefoot{For the other input parameters, see \cref{tab:joint_ret_params}.}
\label{tab:params_cc}
\end{table}

\begin{figure*}[!htb]
\centering
\includegraphics[scale=1.02]{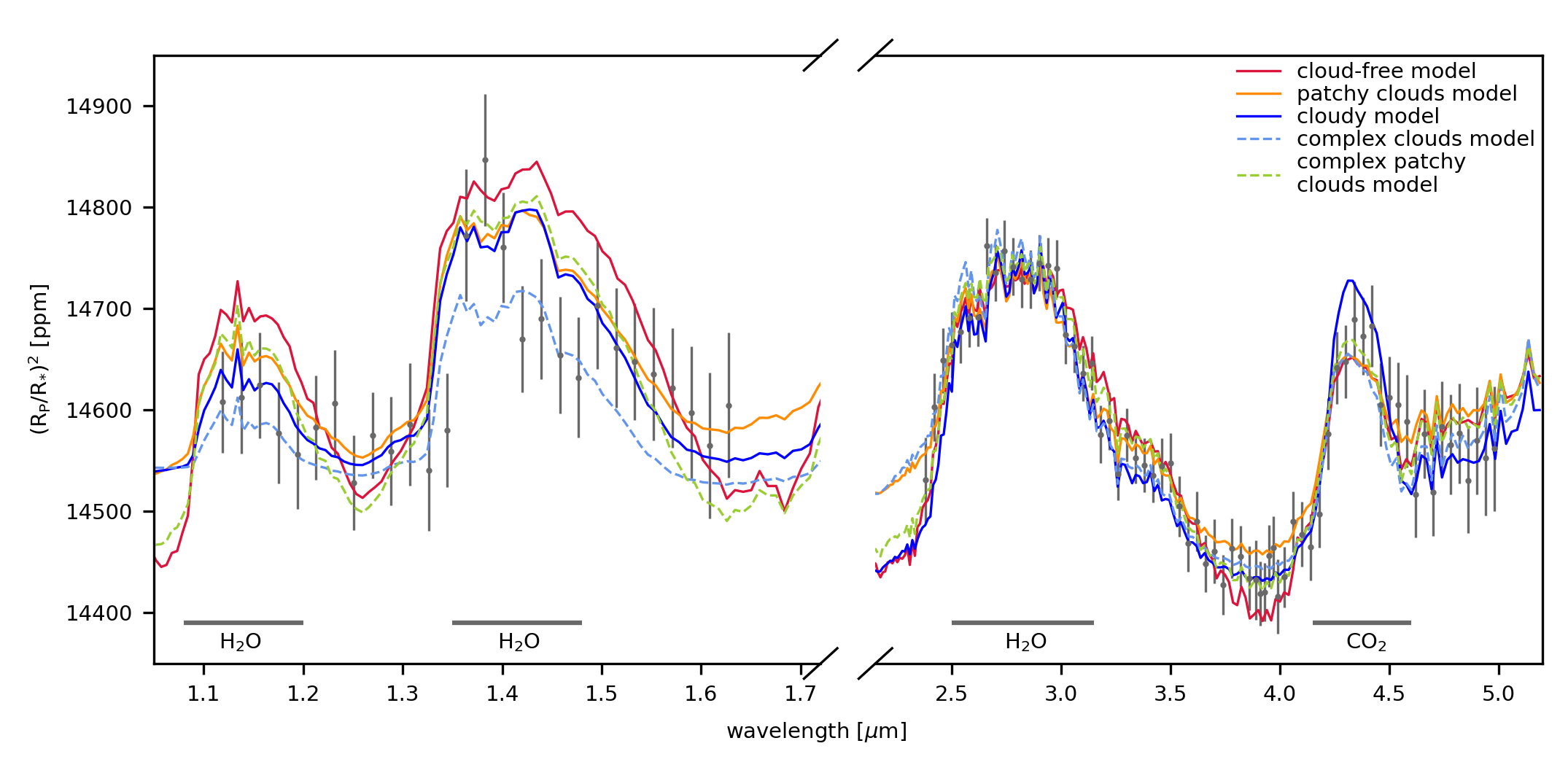}
\vspace*{-10mm}
\caption{Comparison of best-fit spectra for retrievals of the joint HST-JWST spectrum using different cloud models. The fits include all tested absorbing species except for $\mathrm{SO_2}$ and $\mathrm{H_2S}$ (see \cref{tab:joint_ret_results}). The red spectrum is cloud-free, the blue one cloudy, and the orange spectrum is the best fit for patchy clouds. The dashed spectra are best fits using the more complex cloud model described in \cref{subsec:complex_clouds}. We also fit for an offset between the HST and JWST data, which is visible by eye in the figure (for details see \cref{tab:joint_ret_complete_results} and \cref{tab:joint_ret_complete_results_2}).}
\label{fig:comb_cloud_model_comparison}
\end{figure*}

\begin{figure*}[!htb]
\centering
\includegraphics[scale=1.02]{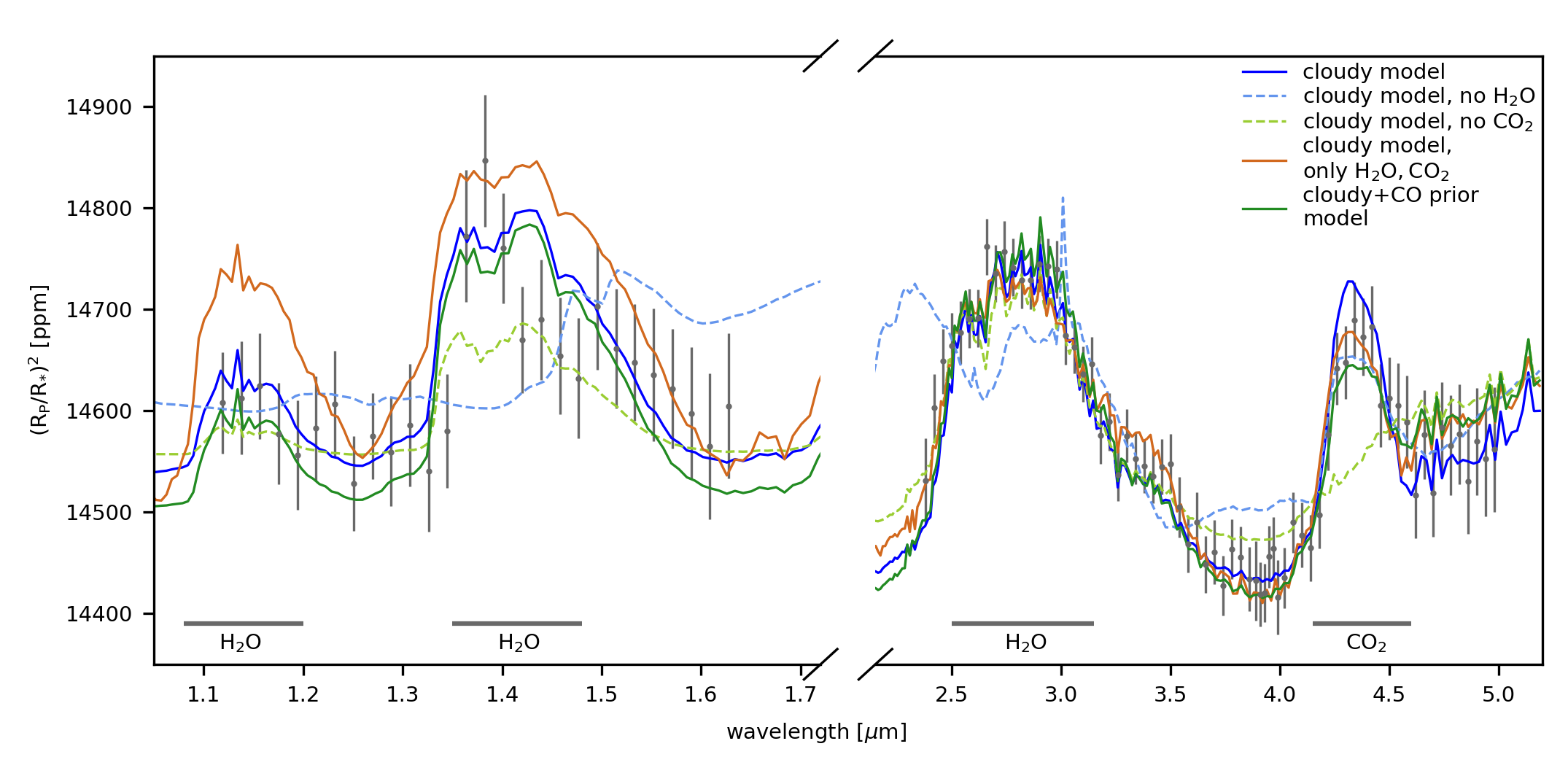}
\vspace*{-10mm}
\caption{Comparison of best-fit spectra for retrievals of the joint HST-JWST spectrum using different molecular absorbers and a grey cloud. The orange spectrum is the best-fit case from the model with the largest Bayesian evidence: a grey cloud and only $\mathrm{H_2O}$ and $\mathrm{CO_2}$ as line absorbers. The light blue dashed spectrum does not include water as an absorber, whereas the light green dashed spectrum is without $\mathrm{CO_2}$. The dark green spectrum is the best fit of the cloudy+CO prior model (see \cref{subsec:metallicities_C/O}). As in \cref{fig:comb_cloud_model_comparison}, there is a vertical offset between the HST and JWST data.}
\label{fig:comb_species_comparison}
\end{figure*}

The results of the retrievals using this more complex cloud model are shown in \cref{tab:joint_ret_results}. All of these models exhibit slightly lower Bayesian evidence than the simple grey cloud-deck model, despite having more free parameters to adjust. Therefore, the more complex model is not necessary to describe the clouds of \planet. Complex cloud and complex patchy cloud models are not significantly ruled out compared to the simple cloud model, with significance levels of $\SI{2.1}{\sigma}$ and $\SI{1.6}{\sigma}$, respectively. On the other hand, the more complex cloud model that includes only $\mathrm{H_2O}$ and $\mathrm{CO_2}$ as line absorbers is ruled out by $\SI{3.6}{\sigma}$ in comparison to the simple cloud model that includes only $\mathrm{H_2O}$ and $\mathrm{CO_2}$. Notably, the retrievals with the more complex cloud model yield higher water abundances. This can be explained by the fact that for the more complex model, the cloud opacity extends above the cloud base at $P_{\rm base}$, and more obscures the water. For the simple cloud model, everything below the cloud location at $P_{\rm cloud}$ is obscured, but the atmosphere is clear above it. Figure~\ref{fig:comb_cloud_model_comparison} also shows the best-fit spectra with the more complex cloud model, compared to those using the simple cloud approximation of a grey opacity at a retrieved pressure. The results for all parameters and retrievals can be found in \cref{tab:joint_ret_complete_results} and \cref{tab:joint_ret_complete_results_2}.

\subsection{Results of the Bayesian model averaging}
\label{subsec:BMA_joint_ret}
\begin{figure*}[!ht]
\centering
     \begin{subfigure}[b]{0.497\textwidth}
         \centering
         \includegraphics[width=1.0\linewidth]{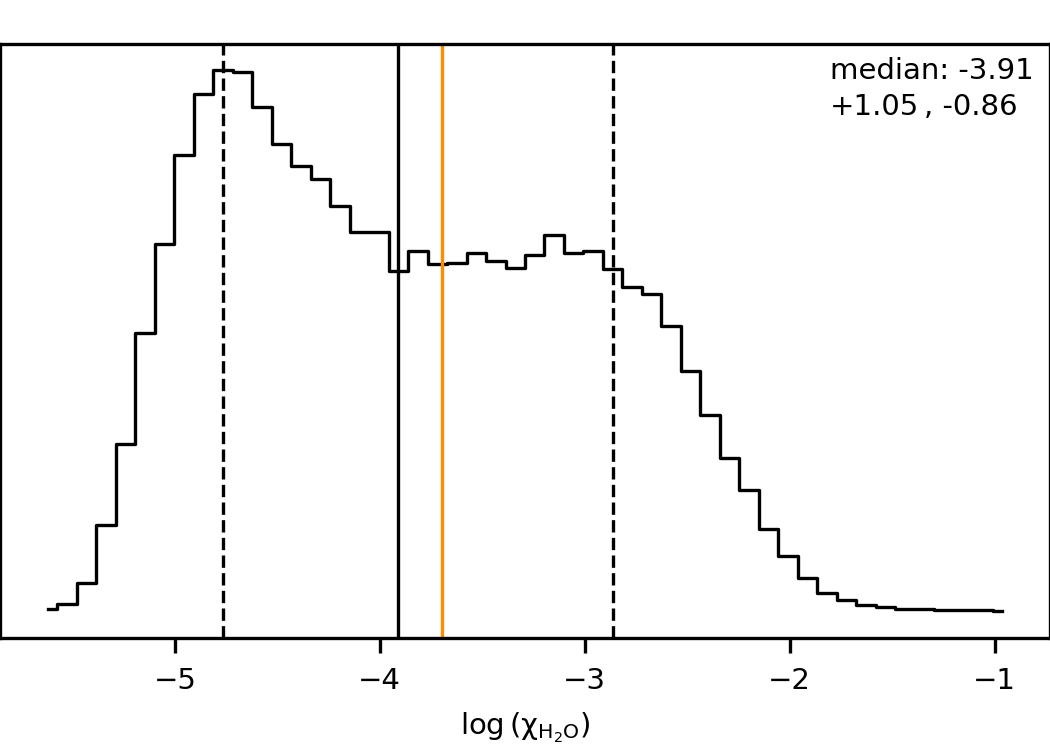}
         \label{fig:BMA_H2O}
     \end{subfigure}
     \hfill
     \begin{subfigure}[b]{0.497\textwidth}
         \centering
         \includegraphics[width=1.0\linewidth]{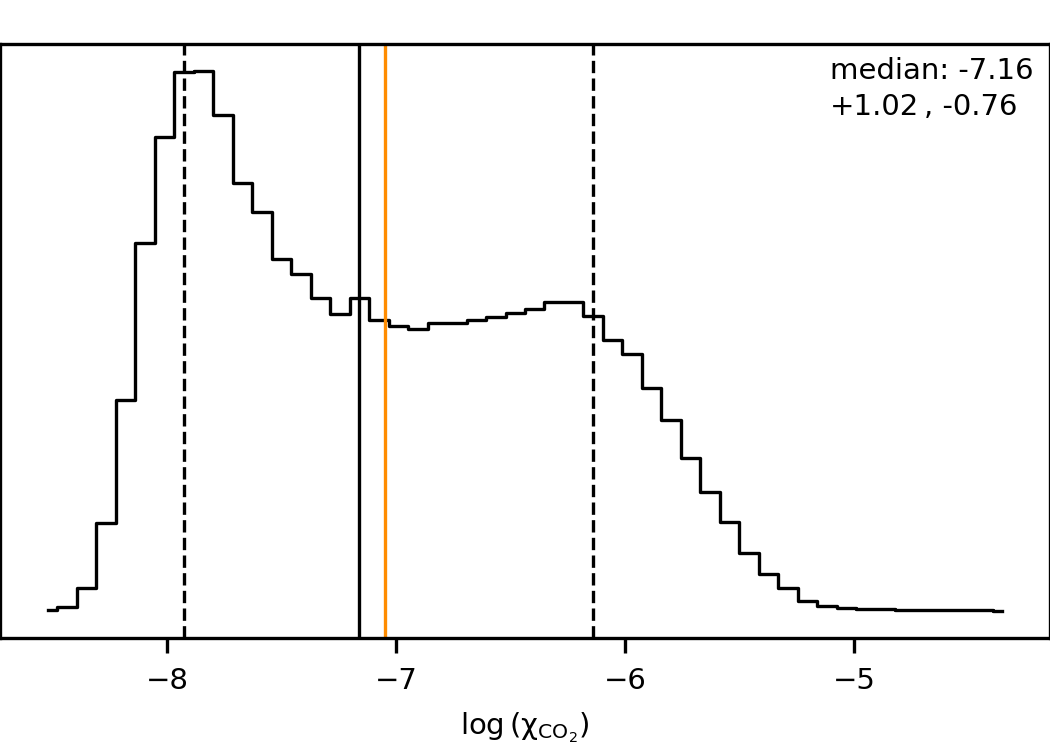}
         \label{fig:BMA_CO2}
     \end{subfigure}
\vspace*{-10mm}
\caption{Bayesian model averaged posterior distributions for $\mathrm{H_2O}$ (left) and $\mathrm{CO_2}$ (right). The black line indicates the median of the distributions, and the dashed lines are the $\pm34.1\%$ confidence regions. The orange lines show the solar value of $\mathrm{log\,(\chi_{H_2O})} = -3.70$ and $\mathrm{log\,(\chi_{CO_2})} = -7.05$ at $\SI{1}{\milli\bar}$ and $\SI{1200}{\kelvin}$.}
\label{fig:BMA_joint_H2O_CO2}
\end{figure*}

As for the HST spectral fits, we used Bayesian model averaging to account for the model uncertainty in the retrieved parameters. To avoid biases in the process, we average the cloud-free, cloudy, patchy clouds, complex clouds, and complex patchy clouds model, excluding the quite similar one species left out cloudy models from the averaging. Since most of the species are not actually detectable, this would lead to a combination of the same cloudy model several times. Figure~\ref{fig:BMA} shows Bayesian model averaged posteriors for the parameters present in most of the models tested. From our uniform cloud models tested, we report a grey cloud deck at a Bayesian model averaged pressure of $P_{\rm cloud} = 1.74^{+4.01}_{-1.58}\cdot10^{-2}$\,bar in the atmosphere of \planet. The averaged water abundance corresponds to a volume mixing ratio of $\mathrm{log(\chi_{H_2O})} = -3.91^{+1.05}_{-0.86}$ (see \cref{fig:BMA_joint_H2O_CO2}). The solar water abundance at a representative pressure level of $\SI{1}{\milli\bar}$ is $-3.70$. Therefore, our measured value agrees with the solar abundance within $\SI{1}{\sigma}$. $\mathrm{CO_2}$ is detected with an abundance of $\mathrm{log(\chi_{CO_2})} = -7.16^{+1.02}_{-0.76}$ (see \cref{fig:BMA_joint_H2O_CO2}), which is also consistent with solar abundances within $\SI{1}{\sigma}$. No other species tested show constrained posterior distributions and therefore we can only provide $\SI{3}{\sigma}$ upper limits on abundances. These are $\mathrm{log(\chi_{CH_4})} = -6.35$, $\mathrm{log(\chi_{NH_3})} = -5.16$, $\mathrm{log(\chi_{HCN})} = -5.99$, $\mathrm{log(\chi_{Na})} = -1.87$, and $\mathrm{log(\chi_{CO})} = -3.26$. Comparing the results of the joint HST-JWST spectrum with the averaged abundances of the different species with the solar values at $\SI{1}{\milli\bar}$, we find that almost all of them (with the exception of CO) are consistent within $\SI{1}{\sigma}$ with the solar values (\cref{fig:BMA}, \cref{fig:BMA_joint_H2O_CO2}) as expected from planet formation models \citep{Öberg_2016, Booth_2017}. This challenges the results of \citet{Madhusudhan_2014} and \citet{MacDonald_2017}, while our results are in agreement with those of \citet{Line_2016}, \citet{Pinhas_2019}, \citet{Welbanks_2019}, \citet{Tsiaras_2018} and \citet{Welbanks_2021}. One caveat is that all of these previous studies based their results on the HST spectrum alone. Our measured upper limits on the abundances of $\mathrm{CH_4}$, $\mathrm{NH_3}$, and $\mathrm{HCN}$ are all lower but still consistent with the corresponding results of \citet{Xue_2023}. Similarly to \cite{Xue_2023}, our findings do not confirm the high-resolution ground-based spectroscopy results of \citet{Giacobbe_2021}, who reported detections of $\mathrm{CH_4}$, $\mathrm{NH_3}$, $\mathrm{HCN}$, and $\mathrm{C_2H_2}$. These findings were also already refuted by \citet{Blain_2024}.

\subsection{Metallicities and C/Os}
\label{subsec:metallicities_C/O}
\begin{figure*}[!htb]
\centering
\includegraphics[width=1.0\linewidth]{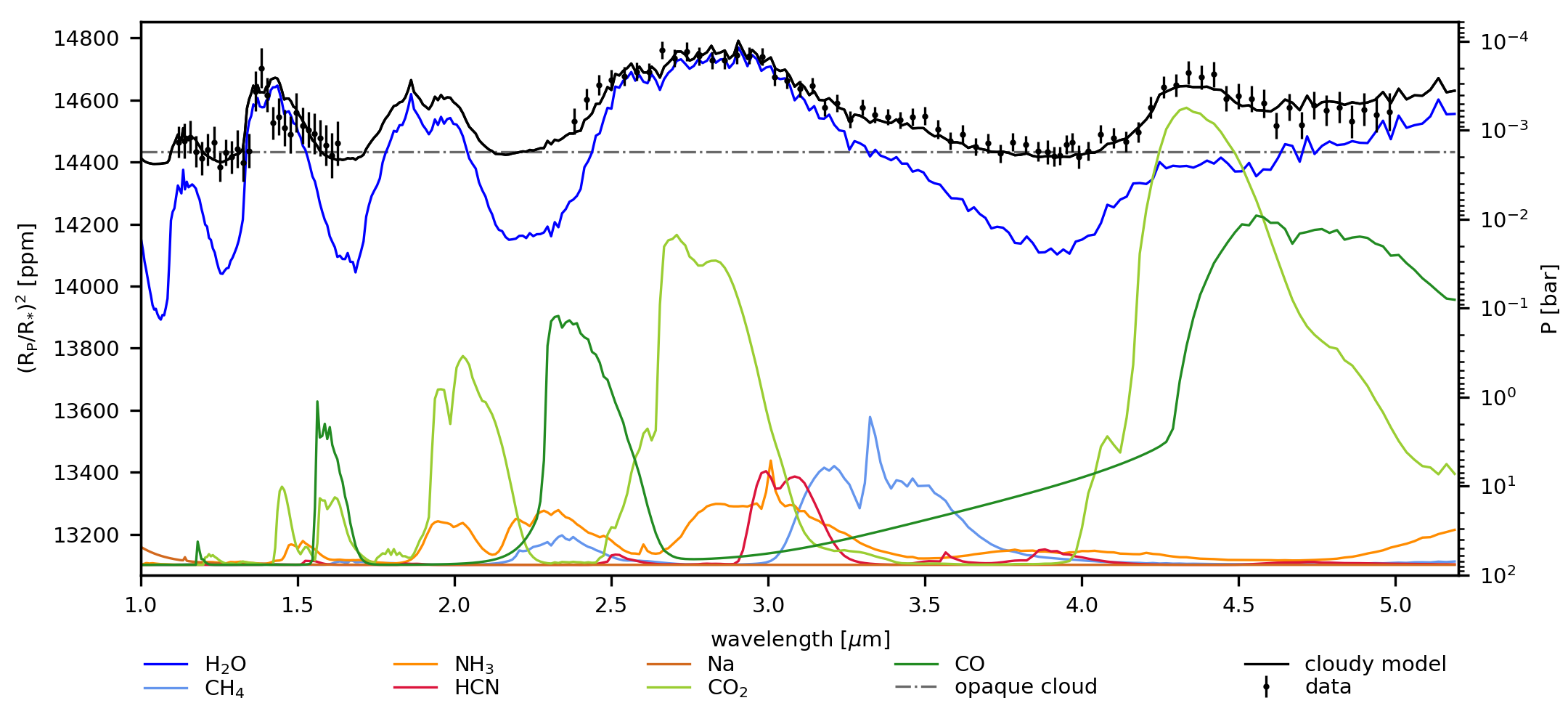}
\vspace{-7mm}
\caption{Contribution of the opacities of line-absorbing species to the transmission spectrum. The best-fit spectrum shows the cloudy+CO prior model (see text). The absorption features of $\mathrm{H_2O}$ and $\mathrm{CO_2}$ are clearly visible, for the latter due to its strong absorption even though the abundance is low (see \cref{fig:BMA_joint_H2O_CO2}).}
\label{fig:opacity_contribution}
\end{figure*}

To compare our free retrieval results with previous work that assumed chemical equilibrium, we calculated metallicity and C/O from the volume mixing ratio posterior distributions of our cloudy retrieval. As we only account for H$_2$, He, and the line-absorbing species tested in the retrievals, the metallicity obtained is a lower estimate of the actual atmospheric metallicity of \planet{} at $\sim\SI{1}{\milli\bar}$. The C/O is calculated from the C/H and O/H ratios we obtained from the retrieved volume mixing ratios. Our results indicate a metal-depleted atmosphere (atmospheric metallicity of $\mathrm{[M/H]} = -1.35^{+1.25}_{-0.73}$), which is consistent with solar values at the $\sim \SI{1}{\sigma}$ level. We also derive an extremely low C/O of $1.33^{+4.79}_{-0.70}\cdot10^{-3}$. The results are likely driven by the low $\mathrm{CO}$ abundance ($\SI{3}{\sigma}$ upper limit of log($\mathrm{\chi_{CO}}$) = -3.26) measured in our spectra and the absence of $\mathrm{CH_4}$ ($\SI{3}{\sigma}$ upper limit of log($\mathrm{\chi_{CH_4}}$) = -6.35). The apparent depletion of $\mathrm{CO}$ relative to the solar value may be an artefact of the clouds, which mute the features in the $\mathrm{CO}$ wavelength band through strong absorption (visible in \cref{fig:opacity_contribution}). Furthermore, \cref{fig:opacity_contribution} clearly shows that in the wavelength range where $\mathrm{CO}$ could be detected ($\SI{4.4}{\micro\meter} - \SI{5.2}{\micro\meter}$) it is additionally obscured by stronger water absorption. Therefore, no features due to $\mathrm{CO}$ absorption are visible. Also, all features from absorbing species other than $\mathrm{H_2O}$ and $\mathrm{CO_2}$ are hidden beneath the cloud deck and consequently are not easy to constrain, despite their opacity still contributing to the total measured opacity. Additionally, the figure illustrates the muting effect of the grey cloud deck on the individual absorbing features, especially those due to water.

\begin{figure*}[!htb]
\centering
\begin{subfigure}{0.5\textwidth}
  \centering
  \includegraphics[width=0.98\textwidth]{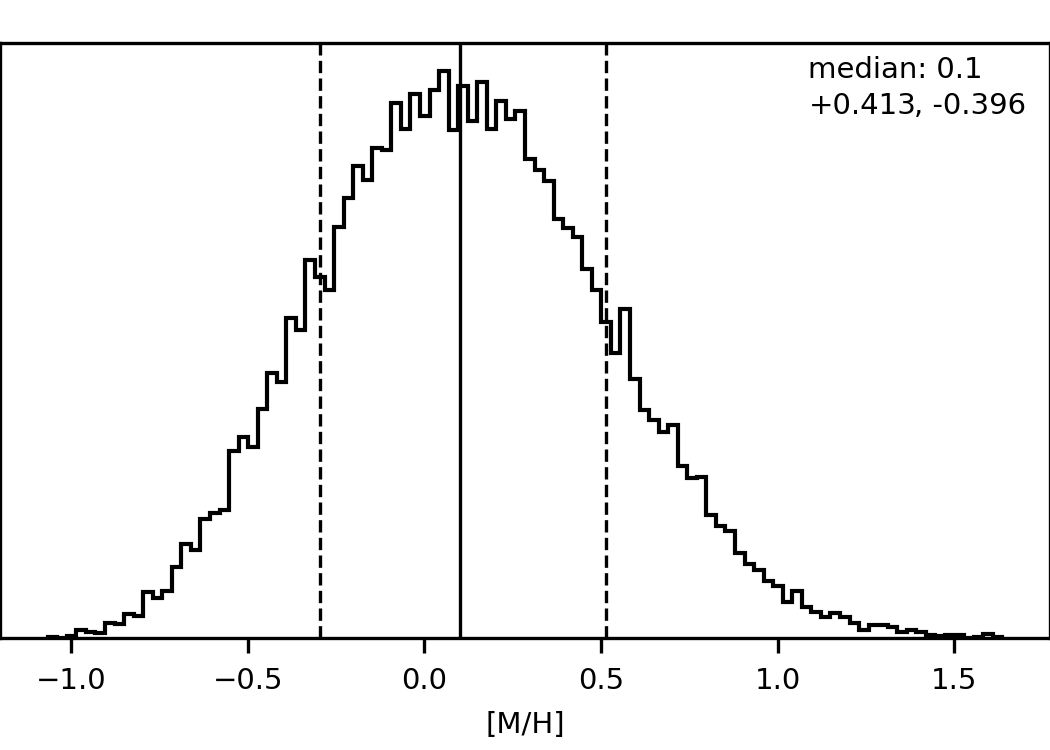}
  \label{fig:metallicity_cloudy}
\end{subfigure}%
\begin{subfigure}{0.5\textwidth}
  \centering
  \includegraphics[width=0.98\textwidth]{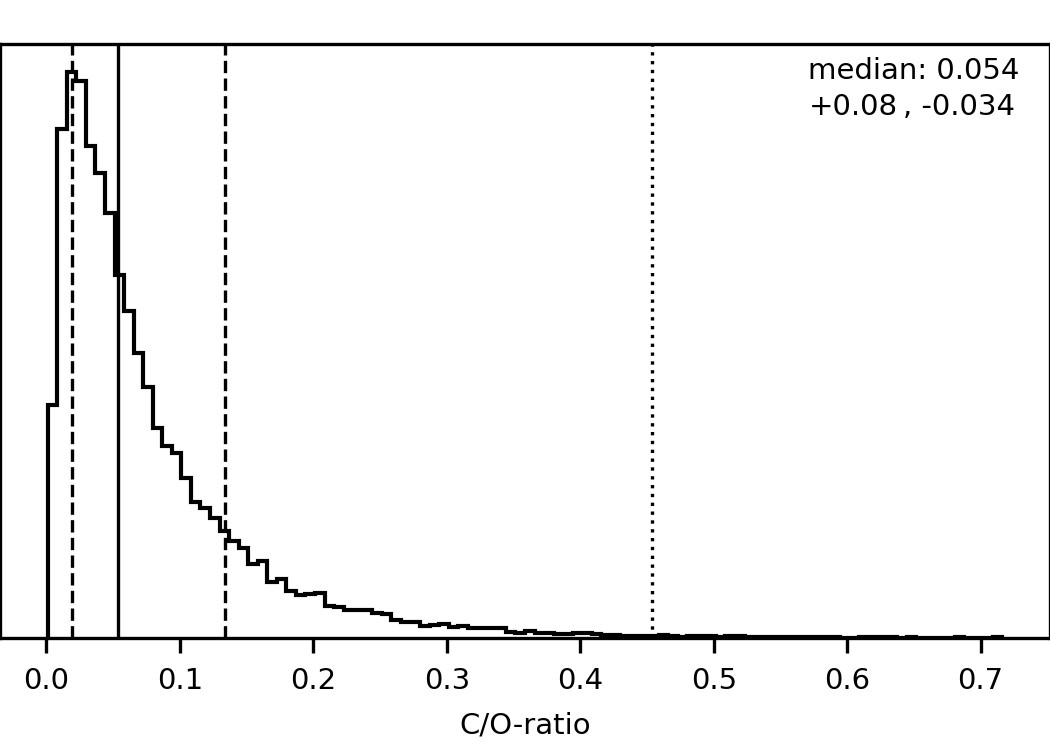}
  \label{fig:C_O_cloudy_high_CO}
\end{subfigure}
 \vspace*{-5mm}
\caption{Metallicity (left) and C/O (right) distribution calculated from the posterior distributions of the cloudy+CO prior retrieval. The black solid lines indicate the median, and the dashed lines depict the $\pm 34.1\%$ confidence regions. The dotted line shows the $\SI{3}{\sigma}$ upper limit of 0.454 for C/O. The same calculations without a tight CO prior yielded a very sub-solar metallicity of $\mathrm{[M/H]} = -1.35^{+1.25}_{-0.73}$ and very low C/O of $1.33^{+4.79}_{-0.70}\cdot10^{-3}$ (see \cref{subsec:metallicities_C/O}).}
\label{fig:metallicity_C_O_cloudy_high_CO}
\end{figure*}

To account for hidden CO, we fit the data again using a ground-based high-resolution $\mathrm{CO}$ abundance measurement from \citet{Brogi_2019} as a prior in a retrieval with a grey cloud deck (uniform prior $U(-3.5, 0.0$) for the CO mass fraction). From here on, this model will be referred to as "cloudy+CO prior". In the ground-based observations $\mathrm{CO}$ is visible, even if we cannot detect it in the JWST data, because for the high-resolution result many individual absorption lines around $\SI{2.3}{\micro\metre}$ are combined for the final detection \citep{Brogi_2019}. Also, at this wavelength, water does not absorb as strongly (cf. \cref{fig:opacity_contribution}). The best-fit spectrum and the corner plot are shown in \cref{fig:comb_species_comparison} and \cref{fig:comb_corner_plot_high_CO}, respectively. The logarithmic Bayesian evidence for this retrieval is $365.4\pm0.2$ and therefore is disfavoured by $\SI{2.6}{\sigma}$ compared to the cloudy retrieval with the broader $\mathrm{CO}$ prior; however, the larger retrieved $\mathrm{CO}$ volume mixing ratio of $\mathrm{log(\chi_{CO})} = -4.15^{+0.51}_{-0.30}$ shifts the metallicity and C/O distribution to more physically reasonable values of $\mathrm{[M/H]} = 0.10^{+0.41}_{-0.40}$ for the metallicity and $0.054^{+0.080}_{-0.034}$ for the C/O (\cref{fig:metallicity_C_O_cloudy_high_CO}). The metallicity is consistent with the solar value and the CO abundance is slightly below solar, but the C/O is significantly subsolar, indicating an atmosphere rich in oxygen and depleted in carbon. 

\subsection{Comparison with previous work}
\label{subsec:comparison_Xue}
In \citet{Xue_2023}, the authors presented results for equilibrium chemistry retrievals of the JWST/NIRCam data (and one combined HST-JWST retrieval), including the planet's radius (at a pressure of $\SI{1}{\bar}$), temperature, metallicity, C/O, the pressure at the cloud top, the cloud fraction, and the mixing ratios of $\mathrm{CH_4}$, $\mathrm{NH_3}$, $\mathrm{C_2H_2}$, and $\mathrm{HCN}$. The only relevant absorbing species they identified were $\mathrm{H_2O}$ and $\mathrm{CO_2}$, with a tentative detection of $\mathrm{H_2S}$. They find a metallicity of $\mathrm{[M/H]} = 0.63^{+0.31}_{-0.24}$ and a C/O of $0.19^{+0.14}_{-0.09}$. Their retrieval favours a patchy cloudy with a covering fraction of around 70\%.

In this work, we focus on free retrievals that do not impose equilibrium chemistry. Our combined HST and JWST free chemistry retrievals yield broadly similar results to \citet{Xue_2023}, with confident detections of H$_2$O, CO$_2$, and clouds. However, even with the CO prior, our inferred metallicity and C/O are still lower than those reported by \citet{Xue_2023}, at $\SI{2.4}{\sigma}$ and $\SI{1.7}{\sigma}$, respectively. One reason for this difference could be that the equilibrium chemistry retrieval of \citet{Xue_2023} is sensitive to the shape of the old, previously published HST spectrum. We find that our results are more consistent with \citet{Xue_2023} when they analyse the JWST data alone. In this case, their retrieved C/O value of $0.08^{+0.09}_{-0.05}$ for this data set alone agrees within $\SI{1}{\sigma}$ with our result.

Some other differences include no evidence for $\mathrm{H_2S}$ in our retrievals; instead, we find a $\SI{3}{\sigma}$ upper limit of $\mathrm{log(\chi_{H_2S})} = -4.00$, due to the unconstrained posterior distribution of the volume mixing ratio. \citet{Xue_2023}'s retrievals suggest a cloud coverage of $\SI{70}{\%}$, whereas our retrievals slightly favour a uniform grey cloud deck. However, the cloud coverage of our patchy cloud retrieval is comparable to their results. A notable feature in their analysis is a correlation between the planet's radius and temperature (see Figure 3 in \citet{Xue_2023}), which our results do not show. The authors suggest that this could be caused by the limited wavelength coverage in their study, which may explain the absence in our analysis that benefits from the broader coverage provided by the HST data. 

During the preparation of this work,  \citet{Verma_2025} also analysed the atmospheric composition of \planet. They find a mostly isothermal atmosphere for \planet{} as well, and obtain within $\SI{1}{\sigma}$ comparable molecular abundances from free chemistry retrievals of the HST/WFC3 and JWST/NIRCam data as ours. In addition, they include HST/STIS observations and note that their slightly lower abundances are sub-solar in all cases. In agreement to our results and the \citet{Xue_2023} study, they only can constrain H$_2$O and CO$_2$ abundances and report upper limits for CH$_4$, NH$_3$, HCN, Na, and CO lower than our results. They however note that their molecular abundances are overestimated when using less data coverage (i.e. only the NIRCam data). In addition \citet{Verma_2025} performed equilibrium chemistry and grid retrievals to constrain the metallicity and C/O. Using only HST/WFC3 and JWST/NIRCam data, they find solar metallicities (in agreement with our results) and subsolar C/O results, comparable but higher than ours. When including the HST/STIS data they report solar C/O values, therefore strongly disagreeing with our results. This they explain by the additional data coverage, especially in the shorter optical wavelengths, which helps to break the degeneracy between cloud and haze effects and molecular abundances, therefore allowing better constraints on the latter. This in turn changes the retrieved metallicity and especially C/O results and is possibly the reason for the disagreement as their retrievals without the HST/STIS data resulted in very sub-solar C/O values as well. 

\subsection{Comparison with \texttt{VULCAN} 1D model}
\label{subsec:compare_VULCAN}
\begin{figure}[!htb]
\centering
\includegraphics[scale=1]{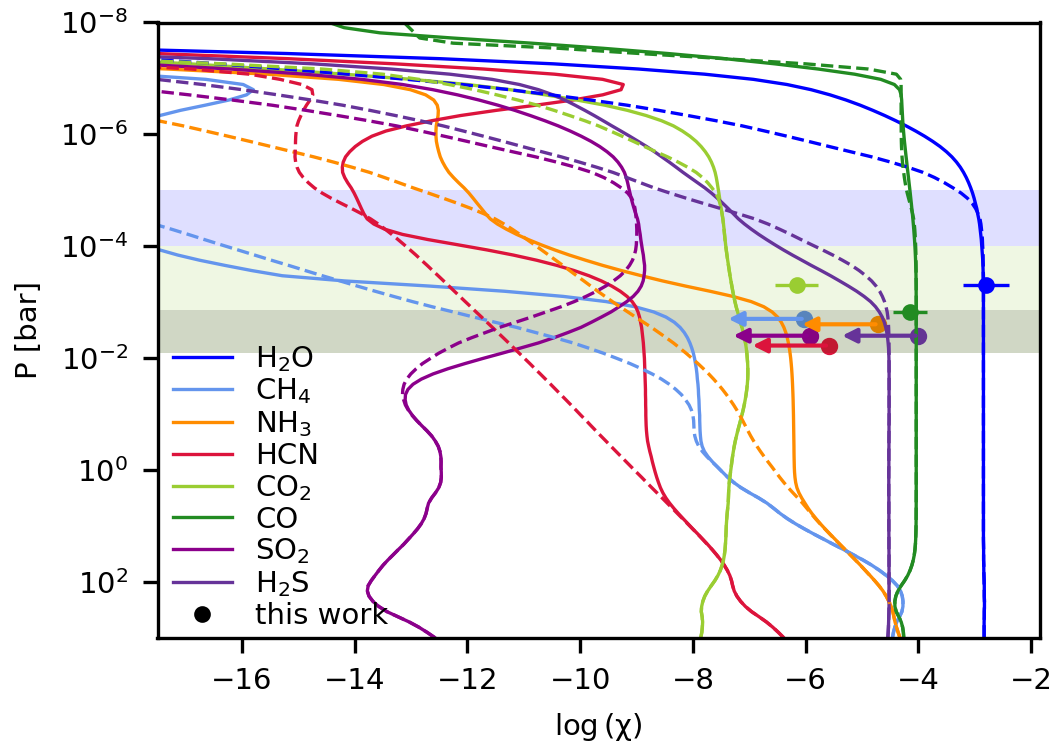}
\vspace*{-2mm}
\caption{Comparison of the cloudy+CO prior retrieval results with the \texttt{VULCAN} 1D photochemical kinetics model by \citet{Tsai_2021} (solid lines) and thermochemical equilibrium abundances (dashed lines). The dots represent the volume mixing ratios from our retrievals (with errorbars for $\mathrm{H_2O}$, $\mathrm{CO_2}$, and $\mathrm{CO}$; upper limits for $\mathrm{CH_4}$, $\mathrm{NH_3}$, $\mathrm{HCN}$, $\mathrm{SO_2}$, and $\mathrm{H_2S}$). The shaded regions represent the atmospheric pressure levels at which the absorption is most active for the species, starting at the cloud level at $\SI{1.44}{\milli\bar}$ (see \cref{fig:opacity_contribution} and \cref{tab:joint_ret_results}). For $\mathrm{CO_2}$ (light green) and $\mathrm{H_2O}$ (light blue), the absorption pressure level reaches higher up in the atmosphere. The Bayesian model averaged results are similar, except for the $\mathrm{H_2O}$, which is smaller than the \texttt{VULCAN} 1D results, and the $\mathrm{CO_2}$ abundance, which agrees with the model prediction.}
\label{fig:compare_VULCAN}
\end{figure}

Finally, we compared our results with predictions from the 1D photochemical \texttt{VULCAN} model \citep{Tsai_2021} (see \cref{fig:compare_VULCAN}). The \texttt{VULCAN} predictions are calculated using a solar metallicity and our caluclated C/O of $= 0.054$. All of our retrieved $\SI{3}{\sigma}$ upper limits are in agreement with the \texttt{VULCAN} model. For the cloudy+CO prior retrieval, only the $\mathrm{CO_2}$ abundance is slightly higher than the \texttt{VULCAN} predictions (cf. \cref{fig:compare_VULCAN}). The differences may arise because our retrievals suggest a much lower C/O than used in the model in \citet{Tsai_2021}. An additional comparison of the Bayesian model averaged results (see \cref{fig:BMA}) with the \texttt{VULCAN} model predictions shows agreeing upper limits for all species (including $\mathrm{CO}$). The $\mathrm{H_2O}$ and $\mathrm{CO_2}$ volume mixing ratios measured from the averaging are lower than those shown in \cref{fig:compare_VULCAN}. The Bayesian model averaged $\mathrm{CO_2}$ abundance is in agreement with the predicted values from the model but the water abundance is lower. 

Figure~\ref{fig:compare_VULCAN} also shows that with our results we cannot distinguish between equilibrium or disequilibrium abundances in \planet's atmosphere. Future ground-based results, which might be able to constrain the abundances of $\mathrm{CH_4}$ or nitrogenous species in the atmosphere, could resolve this ambiguity.

%--------------------------------------------------------------------
\section{Discussion and conclusions}
\label{sec:conclusions}
This study provides a comprehensive analysis of the atmospheric transmission spectrum of the hot Jupiter exoplanet \planet, including an updated fit to the HST/WFC3 data. To measure the abundances of line-absorbing species in the planet's atmosphere, we performed atmospheric retrievals on the re-analysed transmission spectrum with a range of different cloud models. Using only the HST spectrum, the cloud-free model was slightly preferred. The only absorbing species conclusively detected was $\mathrm{H_2O}$, with a sub-solar abundance of $0.71^{+0.59}_{-0.05} \:\times$ solar (Bayesian model averaged value). Although this value is low and in agreement with several previous studies \citep[e.g.][]{Madhusudhan_2014, Barstow_2017, MacDonald_2017}, the $\SI{1}{\sigma}$ uncertainty includes the solar value. Our larger error bars are probably a better representation of the uncertainty in the cloud properties. Our approach to treat the wavelength-dependent systematics in the HST/WFC3 transmission spectrum with special attention did not change the surprisingly low water abundances measured from this spectrum. Different models without clouds yielded similar water abundances (see \cref{tab:ret_list}). Including a grey cloud deck, even if not favoured by this data set but clearly so with the wider wavelength coverage of the added JWST/NIRCam data, pushes the water abundance to more reasonable values and probably describes the state of \planet's atmosphere better, even if it is not statistically favoured. Accounting for the systematics in the Hubble data reduction might therefore change the abundances at a level below the current uncertainties, and other factors (like the cloud coverage) have a more pronounced effect on the results. In general, our study shows that the HST/WFC3 data do not have sufficient wavelength coverage to obtain robust results for \planet.

We also combined our HST/WFC3 data with archival JWST/NIRCam  data and retrieved the atmospheric properties based on the joint spectrum. The combined retrieval resulted in an $\mathrm{H_2O}$ abundance of $0.95^{+0.35}_{-0.17} \:\times$ solar and a $\mathrm{CO_2}$ abundance of $0.94^{+0.16}_{-0.09} \:\times$ solar (with only upper limits for other species). Compared to our fits for the HST data alone, these results are in better agreement with predictions from planet formation models and previous studies by \citet{Pinhas_2019}, \citet{Welbanks_2019}, and \citet{Welbanks_2021}. To directly compare our results with the equilibrium chemistry retrieval results from \citet{Xue_2023}, we calculated the metallicity and C/O from the cloudy retrieval results. In our initial analysis, both values were very low ($\mathrm{[M/H]} = -1.35^{+1.25}_{-0.73}$, C/O = $1.33^{+4.79}_{-0.70}\cdot10^{-3}$), primarily due to the extremely depleted $\mathrm{CO}$ abundance we retrieved, which disagrees with previous ground-based high-resolution measurements of $\mathrm{CO}$ \citep{Brogi_2019}. When we used a more informed prior on the $\mathrm{CO}$ abundance, based on the ground-based high-resolution results, we obtained results similar to those of \citet{Xue_2023}. For this case, the highest likelihood model fits almost as well as the one that used a wider prior (Bayes factor of 3 when the model with the wider prior is used as reference). With this result, we obtained a more realistic, solar-like metallicity of $\mathrm{[M/H]} = 0.10^{+0.41}_{-0.40}$ and a C/O of $0.054^{+0.080}_{-0.034}$ with a $\SI{3}{\sigma}$ upper limit of $0.454$ -- significantly lower than the stellar C/O of $0.47$ \citep{Brewer_2017}.

Our results highlight the sensitivity of fundamental atmospheric parameters -- C/O and metallicity -- to details of the model assumptions. For the HST/WFC3 data alone, we found that cloud-free models were preferred, even though clouds are strongly favoured when the JWST data are added. Similarly, even with both data sets included, CO is not detected, which tends to push the C/O and metallicity towards extremely low values. The fact that CO is not detectable in the JWST spectrum is particularly concerning, given that \planet\ is one of the easiest known transmission spectroscopy targets and CO is strongly detected from the ground \citep{Brogi_2019}. One possible explanation is that at the low spectral resolution of these data, the carbon monoxide features are hidden by strong opacity of clouds and H$_2$O.  Another possibility is that the simplified cloud models used in the retrieval do not fully capture the effect of clouds on the spectrum. To avoid misinterpreting the spectra for this planet and others, a combination of space and high-resolution ground-based data should be considered for other studies as well. Further, our analysis calls into question the use of Bayesian evidence as the sole criterion for model selection. Given that the "best" models missed important features that we know to be present (cloud, CO), Bayesian model averaging is a safer approach that better captures the true model uncertainty. This will be even more important for lower S/N targets, such as sub-Neptunes and Earth-like planets, where the data quality will never match what we have for a (comparatively) easy target like \planet.

To further improve our understanding of \planet's atmosphere, future studies should consider that to get a full census of the atmospheric composition, models also need to account for species with fewer or weaker lines that are not easily visible in low-resolution transmission spectra. In general, the isothermal pressure-temperature profile and the cloud assumptions used here may also be too simplistic to fully reveal the atmospheric composition and structure of \planet. Observing the planet in the mid- to far-infrared wavelength range could provide insights into the composition of the clouds in its atmosphere. Additionally, considering different particle sizes in the clouds could offer a better understanding of their influence on the spectrum.

The low metallicity and C/O results are consistent with the picture that \planet{} formed by core accretion beyond and migrated inward of the snowlines of carbon-bearing species \citep{Öberg_2011}, and experienced metal enrichment through pebble or planetesimal pollution before the planet reached its present-day location \citep{Espinoza_2017, Brewer_2017}. In general, low planetary C/O results suggest a significant amount of pebble accretion during or after the formation, and a formation further out in the protoplanetary disk, since there planets can accrete more icy bodies, which are in general more oxygen-rich \citep{Cridland_2019}. According to \citep{Penzlin_2024}, very low C/O values are only reached for planets that migrated through the disc from further out, and therefore in addition accreted silicate-rich planetesimals in the inner disk where the carbon is already destroyed. These studies, together with our measured C/O result for \planet{}, may indicate an original formation far from its current location close to the star and strong accretion processes of solids before and during a phase of disk migration. However, these models have a number of uncertainties that make it challenging at this stage to draw firm conclusions regarding the formation of \planet{} \citep{Molliere_2022, Feinstein_2025}.
%--------------------------------------------------------------------
\FloatBarrier

\bibliographystyle{aa}
\bibliography{bibliography}
%--------------------------------------------------------------------
\clearpage
\begin{appendix}
\onecolumn
\section{Additional Figures and Tables} \label{sec:appendix}
\noindent\begin{minipage}{\textwidth}
    \centering
    \includegraphics[scale=0.33]{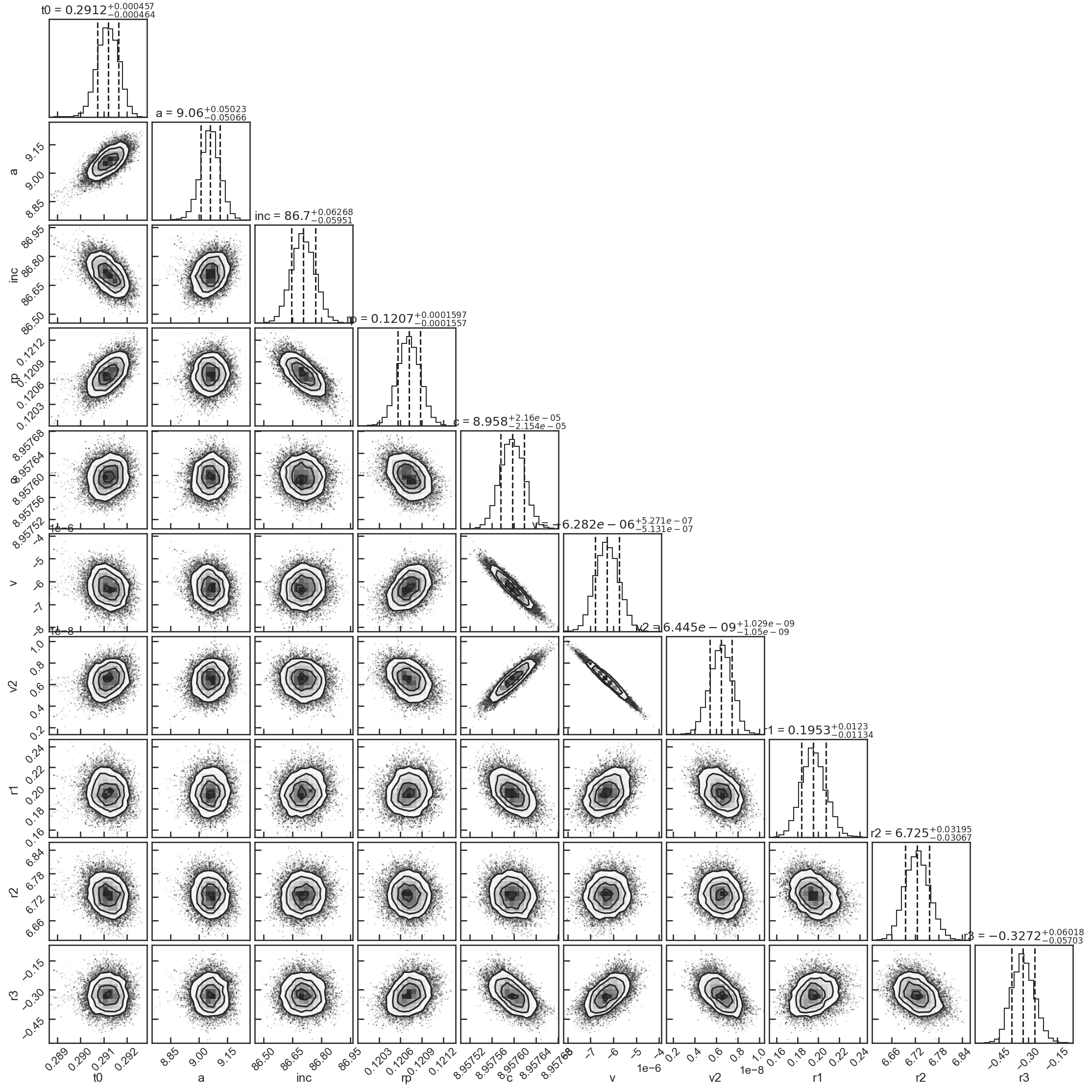}
    \captionof{figure}{Corner plot with 1, 2, 3 $\sigma$ contours for the white light curve fit (\cref{fig:wlc_fit}), including all fit parameters.}
    \label{fig:wlc_corner}
\end{minipage}

\begin{figure*}[htb]
\centering
\includegraphics[scale=1]{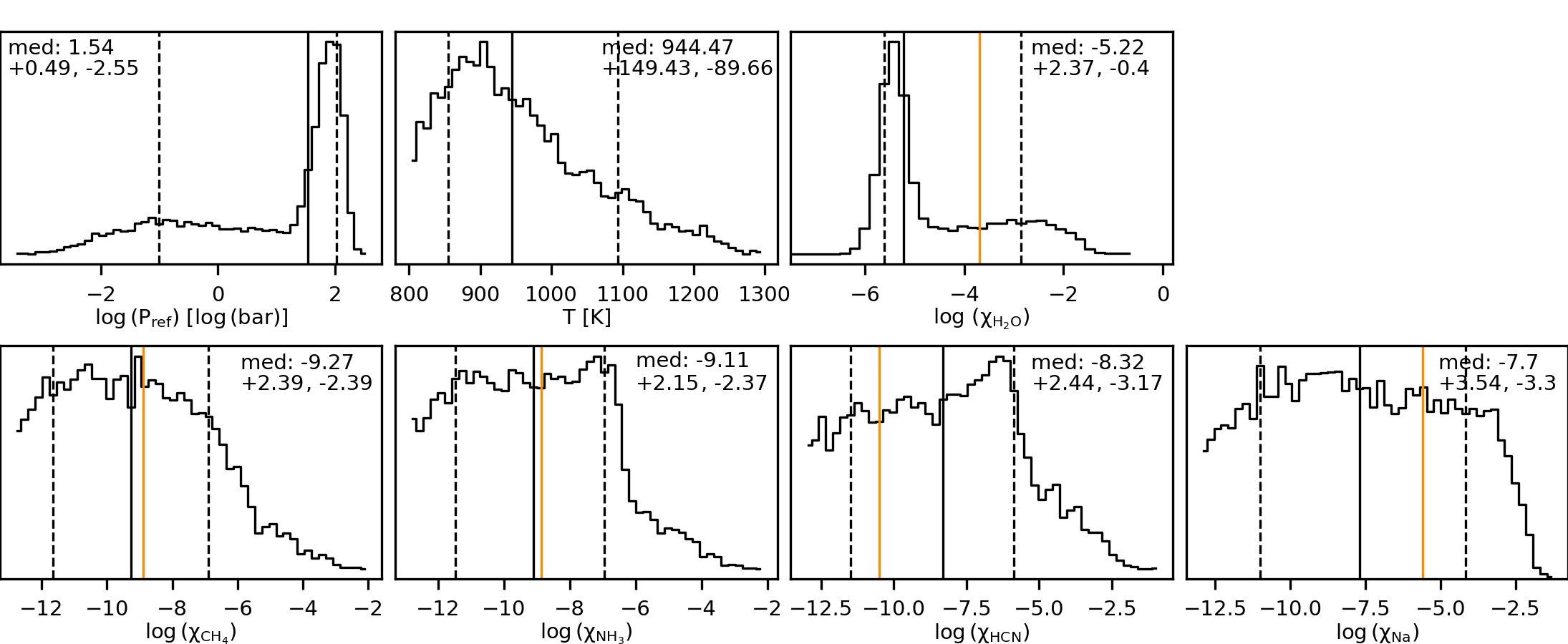}
\caption{Bayesian model averaged results for the all parameters from the "model\_ramp" HST/WFC3 retrievals. The black line indicates the median of the distributions, and the dashed lines show the $\pm34.1\%$ confidence regions. The abundances for the line-absorbing species are given in log volume mixing ratios. The orange lines indicate the solar value at $\SI{1}{\milli\bar}$ and $\SI{1200}{\kelvin}$.}
\label{fig:BMA_HST}
\end{figure*}

\begin{figure*}[!htb]
\centering
\includegraphics[scale=1]{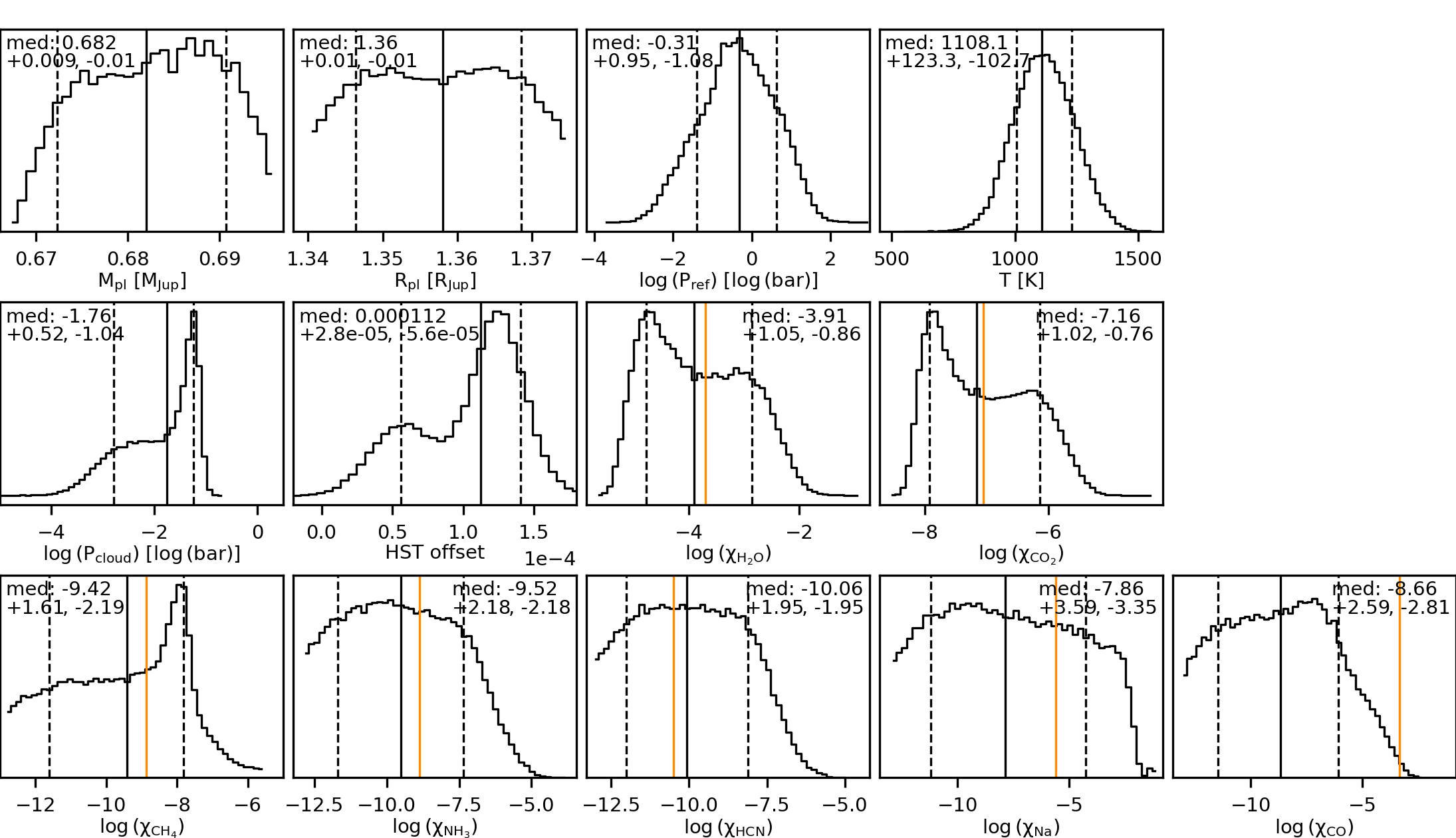}
\caption{Averaged posterior distributions for all parameters from the combined HST-JWST retrievals using BMA. The black line indicates the median of the distributions, and the dashed lines are the $\pm34.1\%$ confidence regions, respectively. The orange lines indicate the solar value at $\SI{1}{\milli\bar}$ and $\SI{1200}{\kelvin}$.}
\label{fig:BMA}
\end{figure*}

\begin{figure*}[!htb]
\centering
\includegraphics[scale=0.41]{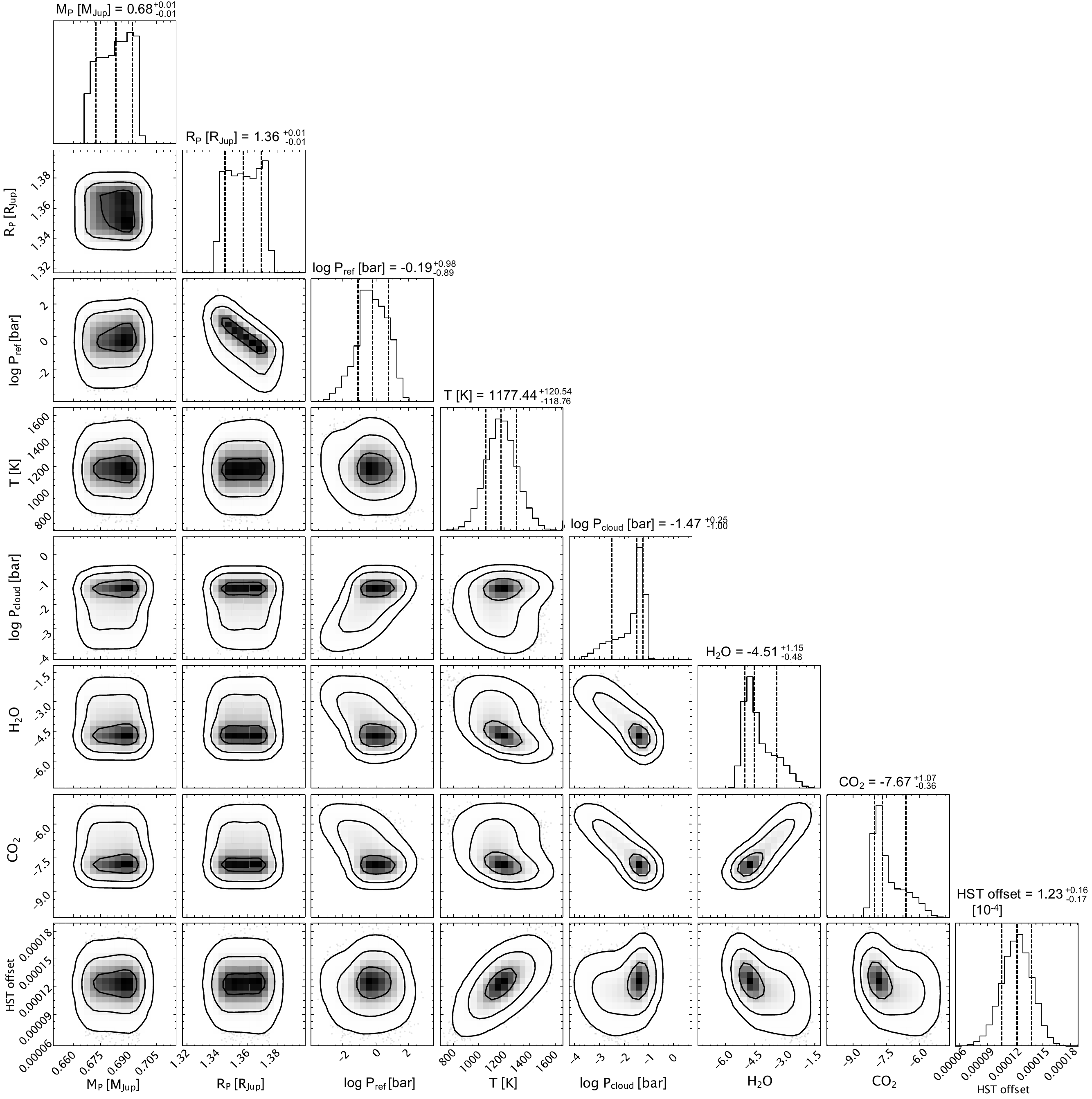}
\caption{Corner plot for the cloudy retrieval of the joint HST-JWST spectrum, only including $\mathrm{H_2O}$ and $\mathrm{CO_2}$ as line absorbers. This is the model with the largest Bayesian evidence.}
\label{fig:comb_corner_plot_best_ret}
\end{figure*}

\begin{figure*}[!htb]
\centering
\includegraphics[scale=0.338]{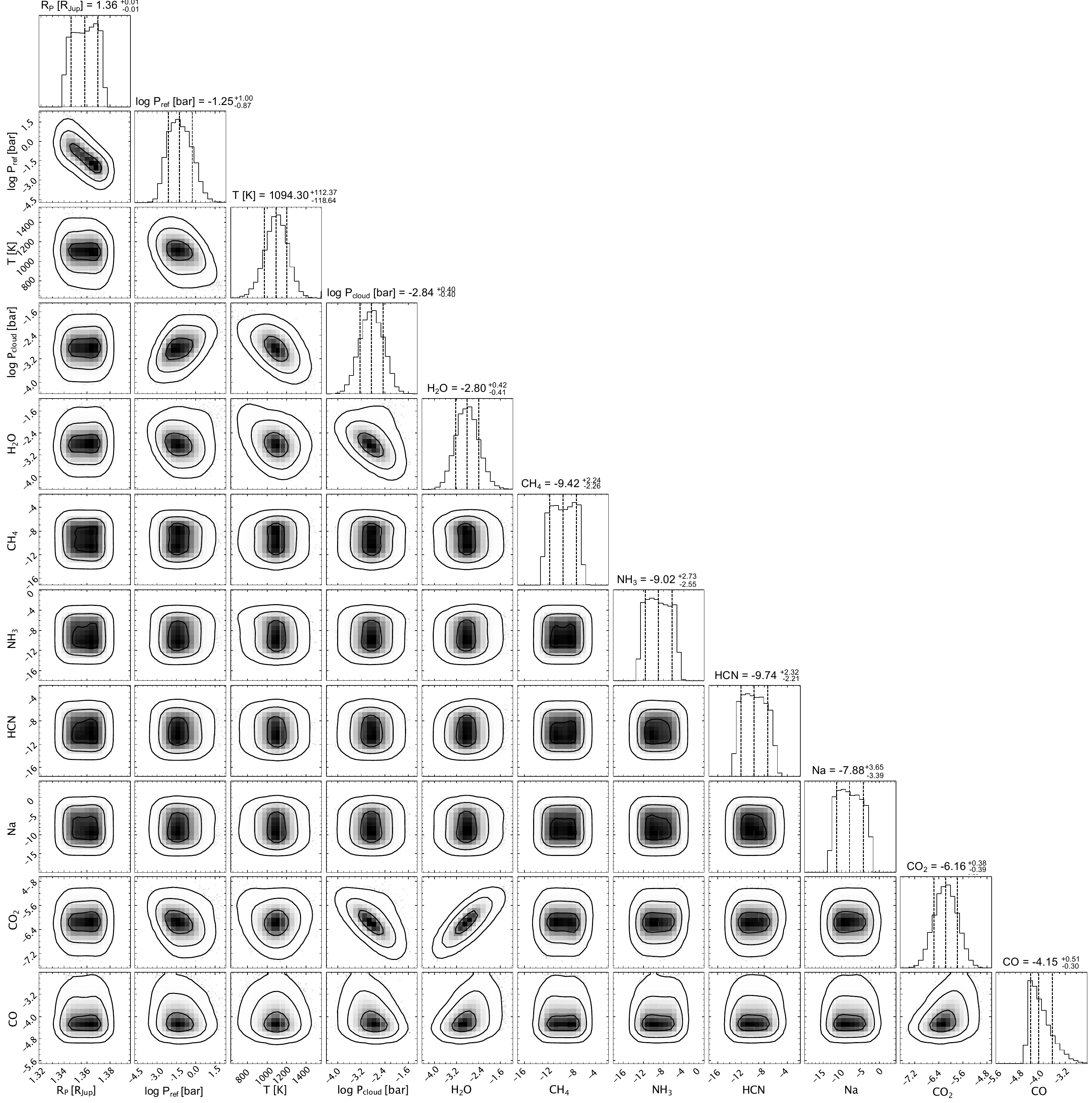}
\caption{Corner plot for the cloudy+CO prior retrieval of the joint HST-JWST spectrum.}
\label{fig:comb_corner_plot_high_CO}
\end{figure*}
\newpage

\renewcommand{\arraystretch}{1.2}
\begin{sidewaystable*}[h]
\caption{Summary of retrievals run on the HST/WFC3 spectra with all the results.}
\centering
\begin{tabular}{>{\centering\arraybackslash}p{2.0cm}>{\centering\arraybackslash}p{1.5cm}|>{\centering\arraybackslash}p{1.5cm}>{\centering\arraybackslash}p{2cm}>{\centering\arraybackslash}p{1.9cm} >{\centering\arraybackslash}p{1.4cm} >{\centering\arraybackslash}p{1.9cm} >{\centering\arraybackslash}p{1.9cm} >{\centering\arraybackslash}p{1.9cm} >{\centering\arraybackslash}p{1.9cm} >{\centering\arraybackslash}p{1.9cm}}
retrieval model & line species & log\,($P_{\rm ref}$) \newline [\SI{}{log\,(\bar)}] &  $T \: [\SI{}{\kelvin}]$ & log\,($P_{\rm cloud}$) \newline [\SI{}{log\,(\bar)}] & cloud \newline fraction & log\,($\chi_{\rm{H_2O}}$) & log\,($\chi_{\rm{CH_4}}$) & log\,($\chi_{\rm{NH_3}}$) & log\,($\chi_{\rm{HCN}}$) & log\,($\chi_{\rm{Na}}$)\\
\hline \hline
"model\_ramp"\newline no clouds & all & $1.88^{+0.22}_{-0.26}$ & $926.46^{+120.59}_{-79.79}$ & \ldots & $-5.50^{+0.26}_{-0.27}$ & $-9.70^{+2.40}_{-2.09}$ & $-9.47^{+2.26}_{-2.29}$ & $-9.16^{+2.46}_{-2.59}$ & $-8.12^{+3.78}_{-3.24}$ & \ldots
\\
"model\_ramp"\newline patchy clouds & all & $0.70^{+1.23}_{-2.31}$ & $963.30^{+158.58}_{-94.52}$ & $-2.88^{+4.06}_{-1.84}$ & $0.63^{+0.17}_{-0.37}$ & $-4.97^{+2.06}_{-0.55}$ & $-9.34^{+2.54}_{-2.19}$ & $-9.11^{+2.37}_{-2.34}$ & $-8.50^{+2.46}_{-2.96}$ & $-7.91^{+3.58}_{-3.24}$
\\
"model\_ramp"\newline clouds & all & $0.63^{+1.33}_{-1.72}$ & $941.27^{+159.06}_{-90.17}$ & $-1.64^{+3.15}_{-1.76}$ & \ldots & $-4.33^{+1.86}_{-1.16}$ & $-8.99^{+2.63}_{-2.51}$ & $-9.04^{+2.39}_{-2.39}$ & $-7.65^{+2.89}_{-3.44}$ & $-7.54^{+3.48}_{-3.25}$
\\
"model\_ramp"\newline no clouds & no $\mathrm{H_2O}$ & $1.78^{+0.25}_{-1.77}$ & $897.02^{+126.46}_{-68.54}$ & \ldots & \ldots & $-5.23^{+4.09}_{-0.26}$ & $-6.28^{+4.29}_{-0.61}$ & $-8.86^{+2.73}_{-2.80}$ & $-7.83^{+3.78}_{-3.40}$ & \ldots
\\
"model\_ramp"\newline no clouds & no $\mathrm{CH_4}$ & $1.93^{+0.22}_{-0.26}$ & $923.52^{+137.94}_{-78.79}$ & \ldots & \ldots & $-5.41^{+0.36}_{-0.27}$ & \ldots & $-9.29^{+2.37}_{-2.37}$ & $-8.83^{+2.37}_{-2.88}$ & $-7.83^{+3.57}_{-3.34}$
\\
"model\_ramp"\newline no clouds & no $\mathrm{NH_3}$ & $1.88^{+0.22}_{-0.26}$ & $930.45^{+117.12}_{-79.57}$ & \ldots & \ldots & $-5.51^{+0.27}_{-0.27}$ & $-9.56^{+2.23}_{-2.19}$ & \ldots & $-9.09^{+2.53}_{-2.67}$ & $-7.90^{+3.62}_{-3.37}$
\\
"model\_ramp"\newline no clouds & no $\mathrm{HCN}$ & $1.87^{+0.22}_{-0.25}$ & $935.45^{+114.29}_{-81.19}$ & \ldots & \ldots & $-5.51^{+0.26}_{-0.28}$ & $-9.64^{+2.33}_{-2.14}$ & $-9.35^{+2.26}_{-2.43}$ & \ldots & $-7.90^{+3.50}_{-3.39}$
\\
"model\_ramp"\newline no clouds & no $\mathrm{Na}$ & $1.89^{+0.22}_{-0.25}$ & $923.10^{+119.35}_{-80.58}$ & \ldots & \ldots & $-5.49^{+0.28}_{-0.27}$ & $-9.65^{+2.24}_{-2.15}$ & $-9.60^{+2.32}_{-2.20}$ & $-8.97^{+2.44}_{-2.67}$ & \ldots
\\
"model\_ramp"\newline no clouds & only $\mathrm{H_2O}$ & $1.84^{+0.25}_{-0.28}$ & $969.17^{+144.21}_{-93.26}$ & \ldots & \ldots & $-5.50^{+0.42}_{-0.27}$ & \ldots & \ldots & \ldots & \ldots
\\
"model\_ramp"\newline clouds & only $\mathrm{H_2O}$ & $1.44^{+0.50}_{-2.36}$ & $979.24^{+165.53}_{-105.96}$ & $-0.62^{+2.42}_{-2.95}$ & \ldots & $-5.12^{+2.95}_{-0.53}$ & \ldots & \ldots & \ldots & \ldots
\\
\textit{mode\_ramp}\newline clouds & no $\mathrm{H_2O}$ & $-0.47^{+1.23}_{-1.12}$ & $956.55^{+184.24}_{-111.14}$ & $-2.80^{+1.25}_{-1.10}$ & \ldots & \ldots & $-3.62^{+1.21}_{-1.47}$ & $-5.89^{+1.90}_{-4.52}$ & $-8.03^{+3.50}_{-3.26}$ & $-8.20^{+3.50}_{-3.10}$ 
\\
\hline
Deming et al.\newline no clouds & all & $1.57^{+0.19}_{-0.21}$ & $981.39^{+146.42}_{-94.26}$ & \ldots & \ldots & $-5.24^{+0.26}_{-0.28}$ & $-9.63^{+2.38}_{-2.18}$ & $-6.64^{+0.42}_{-3.22}$ & $-8.10^{+1.94}_{-3.39}$ & $-8.58^{+3.41}_{-2.92}$
\\
Deming et al.\newline patchy clouds & all & $1.42^{+0.27}_{-0.98}$ & $1015.91^{+148.92}_{-115.86}$ & $-0.16^{+2.04}_{-2.81}$ & $0.44^{+0.32}_{-0.29}$ & $-5.12^{+0.51}_{-0.30}$ & $-9.46^{+2.46}_{-2.20}$ & $-6.53^{+0.55}_{-2.75}$ & $-7.88^{+1.91}_{-3.33}$ & $-8.41^{+3.13}_{-3.01}$
\\
Deming et al.\newline clouds & all & $1.40^{+0.30}_{-1.91}$ & $1037.59^{+160.59}_{-132.08}$ & $0.28^{+1.81}_{-2.65}$ & \ldots & $-5.05^{+1.62}_{-0.37}$ & $-9.06^{+2.60}_{-2.40}$ & $-6.85^{+0.70}_{-3.72}$ & $-6.51^{+2.28}_{-4.27}$ & $-8.23^{+3.24}_{-3.06}$
\\
Deming et al.\newline no clouds & no $\mathrm{H_2O}$ & $-1.61^{+2.80}_{-0.31}$ & $1061.80^{+158.50}_{-195.10}$ & \ldots & \ldots & \ldots & $-1.37^{+0.16}_{-3.14}$ & $-2.09^{+0.16}_{-3.13}$ & $-8.85^{+3.02}_{-2.62}$ & $-8.22^{+3.42}_{-3.00}$
\\
Deming et al.\newline no clouds & only $\mathrm{H_2O}$ & $1.33^{+0.14}_{-0.10}$ & $1173.80^{+70.44}_{-84.98}$ & \ldots & \ldots & $-5.58^{+0.15}_{-0.13}$ & \ldots & \ldots & \ldots & \ldots
\\
Deming et al.\newline clouds & only $\mathrm{H_2O}$ & $1.31^{+0.14}_{-0.12}$ & $1178.93^{+75.96}_{-84.80}$ & $1.12^{+1.27}_{-1.24}$ & \ldots & $-5.57^{+0.16}_{-0.13}$ & \ldots & \ldots & \ldots & \ldots \\
\hline
\label{tab:ret_complete_results}
\end{tabular}
\vspace*{-7mm}
\tablefoot{The abundances of all line-absorbing species are converted to log volume mixing ratios. For a description of the cloud models see \cref{tab:clouds}.}
\end{sidewaystable*}

\newpage

\renewcommand{\arraystretch}{1.5}
\begin{sidewaystable*}[h]
\caption{Summary of retrievals run on the combined HST-JWST spectra.} 
\label{tab:joint_ret_complete_results}
\centering 
\begin{tabular}{>{\centering\arraybackslash}p{1.38cm}|
>{\centering\arraybackslash}p{1.37cm}>{\centering\arraybackslash}p{1.37cm}>{\centering\arraybackslash}p{1.42cm}|>{\centering\arraybackslash}p{1.37cm}>{\centering\arraybackslash}p{1.37cm}>{\centering\arraybackslash}p{1.37cm}>{\centering\arraybackslash}p{1.37cm}>{\centering\arraybackslash}p{1.37cm}>{\centering\arraybackslash}p{1.37cm}>{\centering\arraybackslash}p{1.37cm}>{\centering\arraybackslash}p{1.37cm}>{\centering\arraybackslash}p{1.37cm} >{\centering\arraybackslash}p{1.37cm}}
\hline
retrieval model & no clouds & patchy clouds & clouds & clouds & clouds & clouds & clouds & clouds & clouds & clouds & clouds & clouds & clouds\\
\hline
line species & all & all & all & no $\mathrm{H_2O}$ & no $\mathrm{CH_4}$ & no $\mathrm{NH_3}$ & no $\mathrm{HCN}$ & no $\mathrm{Na}$ & no $\mathrm{CO_2}$ & no $\mathrm{CO}$ & only \newline $\mathrm{H_2O}$, $\mathrm{CO_2}$ & with $\mathrm{SO_2}$ & with $\mathrm{H_2S}$\\ 
\hline \hline 
$M_{\rm pl} \: [\SI{}{M_{Jup}}]$ & $0.68^{+0.01}_{-0.01}$ & $0.68^{+0.01}_{-0.01}$ & $0.68^{+0.01}_{-0.01}$ & $0.68^{+0.01}_{-0.01}$ & $0.68^{+0.01}_{-0.01}$ & $0.68^{+0.01}_{-0.01}$ & $0.68^{+0.01}_{-0.01}$ & $0.68^{+0.01}_{-0.01}$ & $0.68^{+0.01}_{-0.01}$ & $0.68^{+0.01}_{-0.01}$ & $0.68^{+0.01}_{-0.01}$ & $0.68^{+0.01}_{-0.01}$ & $0.68^{+0.01}_{-0.01}$
\\
$R_{\rm pl} \: [\SI{}{R_{Jup}}]$ & $1.37^{+0.00}_{-0.01}$ & $1.36^{+0.01}_{-0.01}$ & $1.36^{+0.01}_{-0.01}$ & $1.36^{+0.01}_{-0.01}$ & $1.36^{+0.01}_{-0.01}$ & $1.36^{+0.01}_{-0.01}$ & $1.36^{+0.01}_{-0.01}$ & $1.36^{+0.01}_{-0.01}$ & $1.36^{+0.01}_{-0.01}$ & $1.36^{+0.01}_{-0.01}$ & $1.36^{+0.01}_{-0.01}$ & $1.36^{+0.01}_{-0.01}$ & $1.36^{+0.01}_{-0.01}$
\\
log\,($P_{\rm ref}$) \newline [\SI{}{log\,(\bar)}] & $0.54^{+1.04}_{-0.53}$ & $-0.32^{+0.94}_{-0.90}$ & $-0.17^{+0.96}_{-0.94}$ & $0.22^{+0.61}_{-0.58}$ & $-0.28^{+1.01}_{-0.99}$ & $-0.15^{+0.96}_{-0.93}$ & $-0.16^{+0.95}_{-0.91}$ & $-0.18^{+0.98}_{-0.98}$ & $-0.39^{+1.37}_{-1.22}$ & $-0.13^{+0.96}_{-0.94}$ & $-0.19^{+0.98}_{-0.89}$ & $-0.27^{+1.01}_{-1.02}$ & $-0.26^{+1.00}_{-1.01}$
\\
$T \: [\SI{e3}{\kelvin}]$ & $1.053^{+0.071}_{-0.076}$ & $1.148^{+0.117}_{-0.124}$ & $1.158^{+0.112}_{-0.113}$ & $1.566^{+0.024}_{-0.047}$ & $1.148^{+0.109}_{-0.113}$ & $1.158^{+0.112}_{-0.110}$ & $1.166^{+0.114}_{-0.112}$ & $1.158^{+0.116}_{-0.114}$ & $1.027^{+0.108}_{-0.342}$ & $1.168^{+0.117}_{-0.114}$ & $1.177^{+0.121}_{-0.119}$ & $1.154^{+0.115}_{-0.113}$ & $1.156^{+0.113}_{-0.112}$
\\
log\,($P_{\rm cloud}$) \newline [\SI{}{log\,(\bar)}] & \ldots & $-2.16^{+0.68}_{-0.78}$ & $-1.45^{+0.27}_{-1.12}$ & $-0.65^{+0.08}_{-0.06}$ & $-1.62^{+0.37}_{-1.09}$ & $-1.43^{+0.25}_{-1.11}$ & $-1.44^{+0.25}_{-1.10}$ & $-1.47^{+0.28}_{-1.15}$ & $-2.19^{+1.92}_{-0.77}$ & $-1.41^{+0.24}_{-1.09}$ & $-1.47^{+0.25}_{-1.00}$ & $-1.53^{+0.34}_{-1.17}$ & $-1.52^{+0.33}_{-1.19}$
\\
cloud \newline fraction & \ldots & $0.75^{+0.13}_{-0.15}$ & \ldots & \ldots & \ldots & \ldots & \ldots & \ldots & \ldots & \ldots & \ldots & \ldots & \ldots
\\
log\,($\chi_{\rm{H_2O}}$) & $-4.39^{+0.24}_{-0.22}$ & $-4.25^{+0.90}_{-0.60}$ & $-4.47^{+1.26}_{-0.50}$ & \ldots & $-4.26^{+1.20}_{-0.60}$ & $-4.51^{+1.26}_{-0.47}$ & $-4.52^{+1.23}_{-0.47}$ & $-4.46^{+1.29}_{-0.52}$ & $-2.58^{+1.70}_{-1.02}$ & $-4.56^{+1.25}_{-0.46}$ & $-4.51^{+1.15}_{-0.48}$ & $-4.35^{+1.27}_{-0.59}$ & $-4.37^{+1.30}_{-0.57}$
\\
log\,($\chi_{\rm{CH_4}}$) & $-10.36^{+1.67}_{-1.45}$ & $-9.52^{+1.56}_{-2.13}$ & $-8.91^{+1.12}_{-2.56}$ & $-10.62^{+1.52}_{-1.50}$ & \ldots & $-8.85^{+1.05}_{-2.58}$ & $-8.87^{+1.07}_{-2.61}$ & $-8.94^{+1.15}_{-2.60}$ & $-10.09^{+1.84}_{-1.73}$ & $-8.94^{+1.12}_{-2.56}$ & \ldots & $-8.96^{+1.20}_{-2.52}$ & $-8.93^{+1.17}_{-2.53}$
\\
log\,($\chi_{\rm{NH_3}}$) & $-9.56^{+2.47}_{-2.02}$ & $-9.65^{+2.25}_{-2.13}$ & $-9.64^{+2.19}_{-2.11}$ & $-6.29^{+0.05}_{-0.05}$ & $-9.61^{+2.24}_{-2.17}$ & \ldots & $-9.60^{+2.17}_{-2.15}$ & $-9.70^{+2.24}_{-2.10}$ & $-9.47^{+2.28}_{-2.14}$ & $-9.75^{+2.19}_{-2.06}$ & \ldots & $-9.59^{+2.19}_{-2.14}$ & $-9.63^{+2.22}_{-2.14}$
\\
log\,($\chi_{\rm{HCN}}$) & $-10.00^{+1.93}_{-1.81}$ & $-10.15^{+1.95}_{-1.91}$ & $-10.14^{+1.96}_{-1.91}$ & $-11.13^{+1.41}_{-1.32}$ & $-10.12^{+2.00}_{-1.93}$ & $-10.16^{+2.02}_{-1.92}$ & \ldots & $-10.16^{+2.02}_{-1.94}$ & $-10.07^{+2.10}_{-1.93}$ & $-10.21^{+1.99}_{-1.91}$ & \ldots & $-10.07^{+1.98}_{-1.96}$ & $-10.09^{+1.99}_{-1.93}$
\\
log\,($\chi_{\rm{Na}}$) & $-1.31^{+0.10}_{-0.16}$ & $-7.60^{+3.81}_{-3.55}$ & $-7.72^{+3.72}_{-3.47}$ & $-8.39^{+3.13}_{-3.08}$ & $-7.68^{+3.60}_{-3.52}$ & $-7.79^{+3.64}_{-3.41}$ & $-7.86^{+3.77}_{-3.41}$ & \ldots & $-7.93^{+3.65}_{-3.26}$ & $-7.74^{+3.71}_{-3.51}$ & \ldots & $-7.69^{+3.73}_{-3.50}$ & $-7.82^{+3.69}_{-3.42}$
\\
log\,($\chi_{\rm{CO}_2}$) & $-7.73^{+0.22}_{-0.21}$ & $-7.43^{+0.82}_{-0.48}$ & $-7.64^{+1.18}_{-0.37}$ & $-8.52^{+0.08}_{-0.08}$ & $-7.47^{+1.14}_{-0.48}$ & $-7.68^{+1.15}_{-0.34}$ & $-7.68^{+1.15}_{-0.36}$ & $-7.64^{+1.20}_{-0.38}$ & \ldots & $-7.70^{+1.15}_{-0.34}$ & $-7.67^{+1.07}_{-0.36}$ & $-7.54^{+1.20}_{-0.46}$ & $-7.56^{+1.23}_{-0.44}$
\\
log\,($\chi_{\rm{CO}}$) & $-9.92^{+1.95}_{-1.87}$ & $-9.50^{+2.39}_{-2.34}$ & $-9.00^{+2.37}_{-2.67}$ & $-10.05^{+1.98}_{-2.05}$ & $-9.01^{+2.48}_{-2.66}$ & $-9.03^{+2.36}_{-2.65}$ & $-9.05^{+2.35}_{-2.65}$ & $-9.07^{+2.44}_{-2.67}$ & $-5.49^{+1.67}_{-4.26}$ & \ldots & \ldots & $-8.94^{+2.44}_{-2.69}$ & $-8.98^{+2.45}_{-2.67}$
\\
log\,($\chi_{\rm{SO_2}}$) & \ldots & \ldots & \ldots & \ldots & \ldots & \ldots & \ldots & \ldots & \ldots & \ldots & \ldots & $-10.36^{+2.09}_{-2.01}$ & \ldots
\\
log\,($\chi_{\rm{H_2S}}$) & \ldots & \ldots & \ldots & \ldots & \ldots & \ldots & \ldots & \ldots & \ldots & \ldots & \ldots & \ldots & $-9.50^{+2.54}_{-2.44}$
\\
offset \newline [$10^{-4}$] & $1.07^{+0.13}_{-0.12}$ & $1.29^{+0.21}_{-0.20}$ & $1.24^{+0.17}_{-0.17}$ & $2.45^{+0.19}_{-0.18}$ & $1.20^{+0.15}_{-0.16}$ & $1.24^{+0.17}_{-0.17}$ & $1.24^{+0.17}_{-0.17}$ & $1.23^{+0.17}_{-0.17}$ & $0.93^{+0.17}_{-0.50}$ & $1.25^{+0.17}_{-0.17}$ & $1.23^{+0.16}_{-0.17}$ & $1.22^{+0.17}_{-0.17}$ & $1.23^{+0.17}_{-0.17}$
\\
\hline
\end{tabular}
\tablefoot{The abundances of all line-absorbing species are converted to log volume mixing ratios. For the more complex cloud model log\,($P_{\rm cloud}$) is the the cloud base pressure log\,($P_{\rm base}$) (see \cref{eq:cloud_opac}). For a description of the cloud models see \cref{tab:clouds}.}
\end{sidewaystable*}

\newpage

\renewcommand{\arraystretch}{1.5}
\begin{table*}[h]
\caption[]{Summary of retrievals run on the combined HST-JWST spectra using the more complex cloud model or a more informed prior for the $\mathrm{CO}$ abundance.}
\label{tab:joint_ret_complete_results_2}
\centering
\begin{tabular}{>{\centering\arraybackslash}p{3.4cm}|>{\centering\arraybackslash}p{2.2cm} >{\centering\arraybackslash}p{2.2cm} >{\centering\arraybackslash}p{2.2cm} >{\centering\arraybackslash}p{2.2cm}}
\hline
retrieval model & complex clouds & complex patchy clouds & complex clouds & clouds \\
\hline
line species & all & all & only \newline $\mathrm{H_2O}$, $\mathrm{CO_2}$ & all, using a more informed prior for $\mathrm{CO}$ \\ 
\hline \hline
$M_{\rm pl} \: [\SI{}{M_{Jup}}]$ & $0.68^{+0.01}_{-0.01}$ & $0.68^{+0.01}_{-0.01}$ & $0.68^{+0.01}_{-0.01}$ & $0.68^{+0.01}_{-0.01}$
\\
$R_{\rm pl} \: [\SI{}{R_{Jup}}]$ & $1.36^{+0.01}_{-0.01}$ & $1.36^{+0.01}_{-0.01}$ & $1.35^{+0.01}_{-0.01}$ & $1.36^{+0.01}_{-0.01}$
\\
log\,($P_{\rm ref}$) [\SI{}{log\,(\bar)}] & $-0.70^{+1.02}_{-0.98}$ & $-0.61^{+0.91}_{-0.91}$ & $-0.38^{+0.83}_{-0.95}$ & $-1.25^{+1.00}_{-0.87}$
\\
$T \: [\SI{}{\kelvin}]$ & $1012.61^{+76.46}_{-79.58}$ & $1083.91^{+98.79}_{-103.67}$ & $1008.69^{+79.20}_{-83.96}$ & $1094.30^{+112.37}_{-118.64}$
\\
log\,($P_{\rm cloud}$) [\SI{}{log\,(\bar)}] & $-1.27^{+0.99}_{-0.78}$ & $-1.35^{+0.98}_{-0.85}$ & $-1.17^{+0.95}_{-0.77}$ & $-2.84^{+0.40}_{-0.40}$
\\
log\,($\kappa_{\rm base}$) $\left[\SI{}{log \left(\frac{\cm^2}{\gram}\right)}\right]$ & $12.99^{+4.23}_{-4.94}$ & $13.15^{+4.26}_{-5.13}$ & $13.37^{+3.99}_{-5.29}$ & \ldots
\\
$f_{\rm sed}$ & $5.97^{+1.96}_{-1.79}$ & $5.64^{+2.21}_{-1.83}$ & $6.43^{+1.84}_{-1.77}$ & \ldots
\\
log\,($\lambda_0$) [\SI{}{log \,(\micro\meter)}] & $-0.98^{+0.84}_{-1.00}$ & $-1.06^{+0.93}_{-1.17}$ & $-0.91^{+0.81}_{-1.12}$ & \ldots
\\
$p$ & $4.34^{+1.00}_{-1.31}$ & $4.12^{+1.14}_{-1.36}$ & $4.33^{+1.01}_{-1.51}$ & \ldots
\\
cloud fraction & \ldots & $0.77^{+0.10}_{-0.11}$ & \ldots & \ldots
\\
log\,($\chi_{\rm{H_2O}}$) & $-2.81^{+0.47}_{-0.71}$ & $-3.48^{+0.71}_{-0.73}$ & $-2.86^{+0.51}_{-0.96}$ & $-2.80^{+0.42}_{-0.41}$
\\
log\,($\chi_{\rm{CH_4}}$) & $-9.67^{+1.84}_{-1.87}$ & $-10.01^{+1.7}_{-1.73}$ & \ldots & $-9.42^{+2.24}_{-2.26}$
\\
log\,($\chi_{\rm{NH_3}}$) & $-9.49^{+2.32}_{-2.03}$ & $-9.47^{+2.18}_{-2.09}$ & \ldots & $-9.02^{+2.73}_{-2.55}$
\\
log\,($\chi_{\rm{HCN}}$)  & $-9.84^{+2.01}_{-1.90}$ & $-9.89^{+1.91}_{-1.88}$ & \ldots & $-9.74^{+2.32}_{-2.21}$
\\
log\,($\chi_{\rm{Na}}$) & $-8.22^{+3.40}_{-2.92}$ & $-7.92^{+3.34}_{-3.13}$ & \ldots & $-7.88^{+3.65}_{-3.39}$
\\
log\,($\chi_{\rm{CO_2}}$) & $-6.18^{+0.44}_{-0.66}$ & $-6.68^{+0.63}_{-0.65}$ & $-6.23^{+0.46}_{-0.88}$ & $-6.16^{+0.38}_{-0.39}$
\\
log\,($\chi_{\rm{CO}}$) & $-6.35^{+1.99}_{-3.74}$ & $-8.29^{+2.57}_{-2.86}$ & \ldots & $-4.15^{+0.51}_{0.30}$
\\
offset & $5.24^{+2.31}_{-2.11}\cdot10^{-5}$ & $6.60^{+2.81}_{-2.54}\cdot10^{-5}$ & $5.54^{+2.36}_{-2.03}\cdot10^{-5}$ & $1.12^{+0.16}_{-0.16}\cdot10^{-4}$ 
\\
\hline
\end{tabular}
\tablefoot{The abundances of all line-absorbing species are converted to log volume mixing ratios. For the more complex cloud model, log\,($P_{\rm cloud}$) is the the cloud base pressure log\,($P_{\rm base}$) (see \cref{eq:cloud_opac}). For a description of the cloud models see \cref{tab:clouds}.}
\end{table*}
\end{appendix}

\end{document}